\documentclass[twocolumn]{aastex631}

\usepackage{amsmath}
\usepackage{empheq}
\usepackage{graphicx}
\expandafter\let\csname equation*\endcsname\relax
  \expandafter\let\csname endequation*\endcsname\relax 
\usepackage{mathrsfs}
\usepackage{textcomp}
\usepackage{enumitem}   
\usepackage{gensymb}
\usepackage{hyperref}
\usepackage{graphicx}
\usepackage[caption=false]{subfig}
\usepackage{multirow}
\usepackage{longtable}
\usepackage[flushleft]{threeparttable}
\usepackage{booktabs} 
\usepackage{CJK}
\usepackage{verbatim}

\hypersetup{colorlinks, linkcolor={blue}, citecolor={blue}, urlcolor={blue}} 

\definecolor{DarkOrange}{RGB}{204, 85, 0}
\definecolor{LincolnGreen}{RGB}{17, 102, 0}

\usepackage{xspace}

\newcommand\nicer{\textit{NICER}\xspace}

\newcommand\swift{\textit{Swift}\xspace}
\newcommand\xmm{\textit{XMM-Newton}\xspace}

\newcommand\rosat{\textit{ROSAT}\xspace}

\newcommand\hst{\textit{HST}\xspace}
\newcommand\swiflong{\textit{Neil Gehrels Swift Observatory}\xspace}
\newcommand\W {{W^r_{\ \phi}}}

\newcommand\ergs{erg s$^{-1}$\xspace}
\newcommand\fluxunit{erg s$^{-1}$ cm$^{-2}$\xspace}
\newcommand\msun{$M_\odot$\xspace}

\newcommand\target{GSN\,069\xspace}

\begin{document}
\pagenumbering{arabic}
\defcitealias{Guolo2025}{Paper I}

\title{A Time-Dependent Solution for GSN\,069 Disk Evolution and the Nature of Long-Lived Tidal Disruption Events}
\author[0000-0002-5063-0751]{M. Guolo}
\affiliation{Bloomberg Center for Physics and Astronomy, Johns Hopkins University, 3400 N. Charles St., Baltimore, MD 21218, USA}

\author{A. Mummery}
\affiliation{Oxford Theoretical Physics, Beecroft Building, Clarendon Laboratory, Parks Road, Oxford, OX1 3PU, United Kingdom}

\author[0000-0002-5311-9078]{A. Ingram}
\affiliation{School of Mathematics, Statistics, and Physics, Newcastle University, Newcastle upon Tyne, NE1 7RU, UK}

\author[0000-0002-2555-3192]{M. Nicholl}
\affiliation{Astrophysics Research Centre, School of Mathematics and Physics, Queens University Belfast, Belfast BT7 1NN, UK}

\author[0000-0003-3703-5154]{S. Gezari}
\affiliation{Space Telescope Science Institute, 3700 San Martin Drive, Baltimore, MD 21218, USA}
\affiliation{Bloomberg Center for Physics and Astronomy, Johns Hopkins University, 3400 N. Charles St., Baltimore, MD 21218, USA}

\author[0000-0002-9633-9193]{E. Nathan}
\affiliation{California Institute of Technology, Pasadena, CA 91125, USA}

\begin{abstract}

We present the implementation of a fully \textit{time-dependent} relativistic disk model—based on the light curve fitting package \texttt{FitTeD}—into the X-ray spectral fitting environment, \texttt{pyXspec}. This implementation enables simultaneous fitting of \textit{multi-epoch} and multi-wavelength spectral data, where the only free parameters are those describing the black hole and the initial conditions, while the subsequent evolution is governed by the dynamical equations of an evolving accretion flow. We use it fit seven epochs of X-ray spectra and two epochs of UV spectra of the `long-lived' tidal disruption event (TDE) and quasi-periodic eruption (QPE) source \target, from 2010 through late-2019. Our results show that such `long-lived', X-ray-bright TDEs—of which \target is a prime, but not unique, example—can naturally be explained within the same framework as events with shorter-lived X-ray emission, like ASASSN-14li and AT2019dsg. Their distinction lies in the `viscous' timescale parameter—tied to the disk's angular momentum transport efficiency—which should be treated as a free parameter when modeling the disk evolution of transient events. We examine the implications for QPE models by tracking the time evolution of disk properties such as mass surface density and accretion rate. We argue that existing QPE models may not be able to reproduce the observed connection between the presence (2018) or absence (2014) of eruptions and the disk properties. In the context of orbiter-disk collision models, the change in mass surface density appears insufficient to explain the needed variation in the eruption's temperature. The absence of eruptions in \target in 2014 remains a challenge for QPE models.

\end{abstract}
\keywords{
Accretion (14);
High energy astrophysics (739); 
Supermassive black holes (1663);\\
X-ray transient sources (1852); 
Time domain astronomy (2109)
}

\vspace{1em}

\section{Introduction}\label{sec:intro}

A diverse array of transient astronomical sources, such as Galactic X-ray binaries, luminous fast blue optical transients \citep[LFBOTs; e.g.,][]{Prentice2018,Margutti2019,Inkenhaag2023,Migliori2024}, and tidal disruption events \citep[TDEs, e.g.,][]{vanvelzen19_late_time_UV,Saxton_20,Mummery2020,Wen2020,Guolo2024,Mummery2024,Guolo_Mummery2025}, are all known to source (at least) some of their emission directly from evolving accretion flows. However, most analytical modeling of the accretion disks of these sources employ steady-state assumptions \citep[e.g.,][]{Shakura1973, Novikov1973}, while the analysis of multi-epoch data treats each observation individually, with the physical parameters of the system allowed to vary in an unconstrained manner between observations. In the case of TDEs, such steady-state assumption is sometimes, inappropriately, coupled to the assumption that the mass accretion rate follows directly the fall-back rate ($\propto t^{-5/3}$)
. 

In reality, such evolving accretion disks, instead of being in a steady-state or having its accretion rate coupled to the fall-back rate, have their physical parameters coupled between epochs by the dynamical equations of motion which govern accretion flows evolution, which themselves are driven by the constraints of mass and angular momentum conservation \citep{Lynden-Bell1974,Eardley1975,Balbus2017}. 

This time-dependent accretion framework as applied to TDEs \citep[e.g.,][] {Cannizzo1990,Mummery2020} has successfully reproduced, and in fact predicted, key properties of the TDE population, such as: the rate of cooling of the peak disk temperature \citep{Guolo2024,Cannizzaro2021}, the presence of late-time UV/optical plateaus and its luminosity correlation with black hole mass \citep{vanvelzen19_late_time_UV,Mummery2024}, and simultaneous X-ray cooling with UV plateauing \citep{Mummery2020,Mummery2024fitted,Guolo_Mummery2025}. They also, uniquely, reproduce the observed UV-plateau and X-ray luminosity functions \citep{Mummery2021xmax,Guolo2024,Mummery_vanVelzen2024,Grotova2025}.

A key class of sources relevant to this study are the so-called `long-lived' TDE candidates \citep[e.g.,][]{Lin2017,Lin2022,Terashima2012,Ho2012,Lin2017b, Lin2011,Lin2018,Lin2020,Lin2018b,Saxton2011,Miniutti2013,Shu2018}. These X-ray bright transients exhibit typical TDE features—thermal soft X-ray spectra, decaying light curves, and cooling peak disk temperatures—without AGN signatures such as a bright infrared bright torus, hard X-ray corona, or persistent broad-line region. Unlike most X-ray bright TDEs, which fade within months to a few years \citep[e.g.,][]{Cannizzaro2021,Guolo2024}, these sources persist for decade-long time-scale. Their nature remains largely unexplored, especially in light of recent advances in relativistic time-dependent disk theory. Among them, \target is the most extensively studied. Discovered in July 2010 by an \xmm slew survey observation \citep{Saxton2011}, its X-ray flux was hundreds of times higher than previous \rosat upper limits. \swift and \xmm monitoring revealed a gradual flux decline and spectral cooling over the years \citep{Miniutti2013,Miniutti2023_longterm}, interrupted in 2020 by a rebrightening event coinciding with an increase in peak disk temperature \citep{Miniutti2023_longterm}.  

In late 2018 and early 2019, \xmm detected high-amplitude X-ray flares from \target, previously absent in 2014, with intensities rising by up to two orders of magnitude in the hardest bands. These quasi-periodic eruptions \citep[QPEs;][]{Miniutti2019}, which in \target recur every $\sim$9 hours, have since been observed in $\sim$10 sources \citep[e.g.,][]{Giustini2020, Arcodia2021, Arcodia2024a, Nicholl2024,Chakraborty25}, with a wide range of recurrence times.
The origin of these QPEs are highly debated — most models can be divided between accretion disk instabilities \citep[e.g.,][]{2022ApJ...928L..18P, Pan2023, Sniegowska2023, Kaur2023} or interactions with a compact orbiting object \citep[e.g.,][]{Zhao2022, King2022, Lu2023, Linial_Sari2023, Linial2023, Franchini2023,Yao2024,Vurm2024}.  

Most of these models link QPEs to the properties of the accretion disk - which powers persistent soft X-ray emission between eruptions. UV/optical counterparts to the disk have now also been detected in four of the sources \citep{Nicholl2024,Guolo2025,Wevers2025,Chakraborty25}. 
In the case of AT2019qiz, the application of time-dependent disk modeling on the TDE light curve, was supportive of orbiter-disk collision models \citep[in finding the disk was sufficiently radially extended to intercept an orbiting body on the relevant orbit][]{Nicholl2024}.
However, multi-epoch X-ray and UV analysis performed in \citet[][hereafter \citetalias{Guolo2025}]{Guolo2025} suggests current models struggle to explain the observed properties of \target’s eruptions in detail, particularly given its inferred outer disk radius and accretion rate. A key technical limitation of \citetalias{Guolo2025} was its use of time-independent approaches, which prevented direct estimates of disk mass and surface density ($\Sigma_{\rm disk}$)—crucial for orbiter-disk interaction models \citep[e.g.,][]{Linial2023,Franchini2023}. Constraining $\Sigma_{\rm disk}$ and its evolution requires a fully time-dependent disk model.

To our knowledge, the only publicly available time-dependent disk fitting tool is \textit{Fitting Transients with Discs} \citep[\texttt{FitTeD},][]{Mummery2024fitted}, already widely applied to TDEs/QPEs \citep[e.g.,][]{Mummery2020,Goodwin2022,Mummery2024,Nicholl2024,Goodwin2024, Chakraborty25}, and LFBOTs \citep{Inkenhaag2023}. However, \texttt{FitTeD} primarily models light curves rather than X-ray spectra. A key goal of this work is to extend \texttt{FitTeD} to X-ray spectroscopy by integrating it with \texttt{pyXspec} \citep[the \texttt{Python} version of \texttt{XSPEC};][]{Arnaud_96}. This should enable tighter constraints on disk and black hole parameters by leveraging X-ray spectral information to break degeneracies—particularly crucial for sources like \target, dominated by X-ray data. To our knowledge, this is the first implementation of a time-dependent model in any X-ray spectral fitting package.

This paper is organized as follows: \S\ref{sec:theory} reviews time-dependent relativistic disk theory; \S\ref{sec:data} introduces the integration of \texttt{FitTeD} into \texttt{pyXspec}, along with data and fitting procedures. Results for \target are presented in \S\ref{sec:results}, followed by a discussion on the model and data constraints (\S\ref{sec:limitations}), highlighting both uncertainties and the robustness of our inferences, while implications for QPE models are presented in \S\ref{sec:qpes}, a brief discussion on the 2020-2022 flare is presented is \S\ref{sec:2020}. The slow evolution of \target, as compared to other TDEs, is discussed in \S\ref{sec:long-lived}, alongside a physical discussion on the nature of ‘long-lived’ TDEs. Conclusions are given in \S\ref{sec:conclusion}.

We assume a $\Lambda$CDM cosmology with $H_0=73\,{\rm km\,s^{-1}\,Mpc^{-1}}$ \citep{Riess2022} and adopt a Bayesian framework, reporting inferred parameters as posterior medians with 68\% credible intervals unless stated otherwise \citep[see][for relevant statistical discussion]{Andrae2010,Buchner2014,Buchner2023}.

\section{Time-Dependent Relativistic Disk Theory and the \texttt{F\lowercase{it}T\lowercase{e}D} package} \label{sec:theory}
The time evolution of flows accreting around a black hole can be understood in terms of the azimuthally-averaged, height-integrated surface density of the flow, \( \Sigma(r, t) \). Classically, the familiar \citet{Lynden-Bell1974} disk evolution equation \citep[e.g.,][]{Frank2002} describes the ``viscous'' (in reality turbulent) evolution of the surface density as a diffusion problem, assuming the disk can be well approximated by a height-integrated form.

The fully relativistic version of this problem is also well understood \citep{Eardley1975,Balbus2017}, where the disk is assumed to evolve in the Kerr metric, which describes the spacetime around the central black hole, characterized by mass \( M_\bullet \) and angular momentum \( J \). The problem is most naturally expressed in cylindrical Boyer-Lindquist coordinates near the equatorial plane, where we define \( r \) (cylindrical radius), \( \phi \) (azimuth), \( z \) (height above the equator), \( t \) (coordinate time at infinity), and \( \text{d}\tau \) (invariant line element). The radius of the innermost stable circular orbit (ISCO) is denoted as \( r_I \), the gravitational radius as \( r_g = GM_\bullet / c^2 \), the spin length parameter as \( a = J / M_\bullet c \), and the dimensionless spin as \( a_\bullet = a / r_g \).  

Assuming the fluid follows circular orbits to leading order, the four-velocity \( U^\mu \) is given by \( U^\mu = \text{d}x^\mu / \text{d}\tau \), with nonzero angular (\( U^\phi \)) and temporal (\( U^0 \)) components. As is well established, a weak magnetic field threading an ionized plasma on circular orbits will trigger the magnetorotational instability \citep[MRI;][]{BalbusHawley91}, leading to turbulence and angular momentum transport. This turbulence is modeled by an anomalous stress tensor \( W^\mu_\nu \), representing the correlation between fluctuations in \( U^\mu \) and \( U_\nu \), or with magnetic fields \citep{Balbus2017}.

The evolution equation for the surface density follows from solving the coupled equations of mass and angular momentum conservation in the flow. Under these assumptions, the governing equation for disk evolution can be written as \citep{Eardley1975,Balbus2017}:
\begin{equation}\label{eq:fund}
{\partial \Sigma \over \partial  t} =  {1\over r U^0} {\partial \ \over \partial  r}\left({U^0\over U'_\phi}    {\partial   \over \partial  r} \left({r\Sigma  \W \over U^0}\right)\right),
\end{equation}
where the primed notation denotes a radial gradient. In the Newtonian limit (\( r \gg r_{g} \), \( U^0 \rightarrow 1 \)), this equation reduces to the familiar \citet{Lynden-Bell1974} disk evolution equation, e.g., \citet{Frank2002}. 

To solve this equation, the functional forms of the four-velocity components for circular orbits in the equatorial plane of the Kerr metric are required, which we reproduce in Appendix \ref{app:math} for completeness. It is evident from Eq.~\ref{eq:fund} and Appendix \ref{app:math} that the inclusion of general relativistic effects significantly increases the algebraic complexity of the governing disk equation compared to the classical formulation \citep[e.g., compare to][]{Lynden-Bell1974,Frank2002}.

\begin{figure}[h!]
	\centering
	\includegraphics[width=\columnwidth]{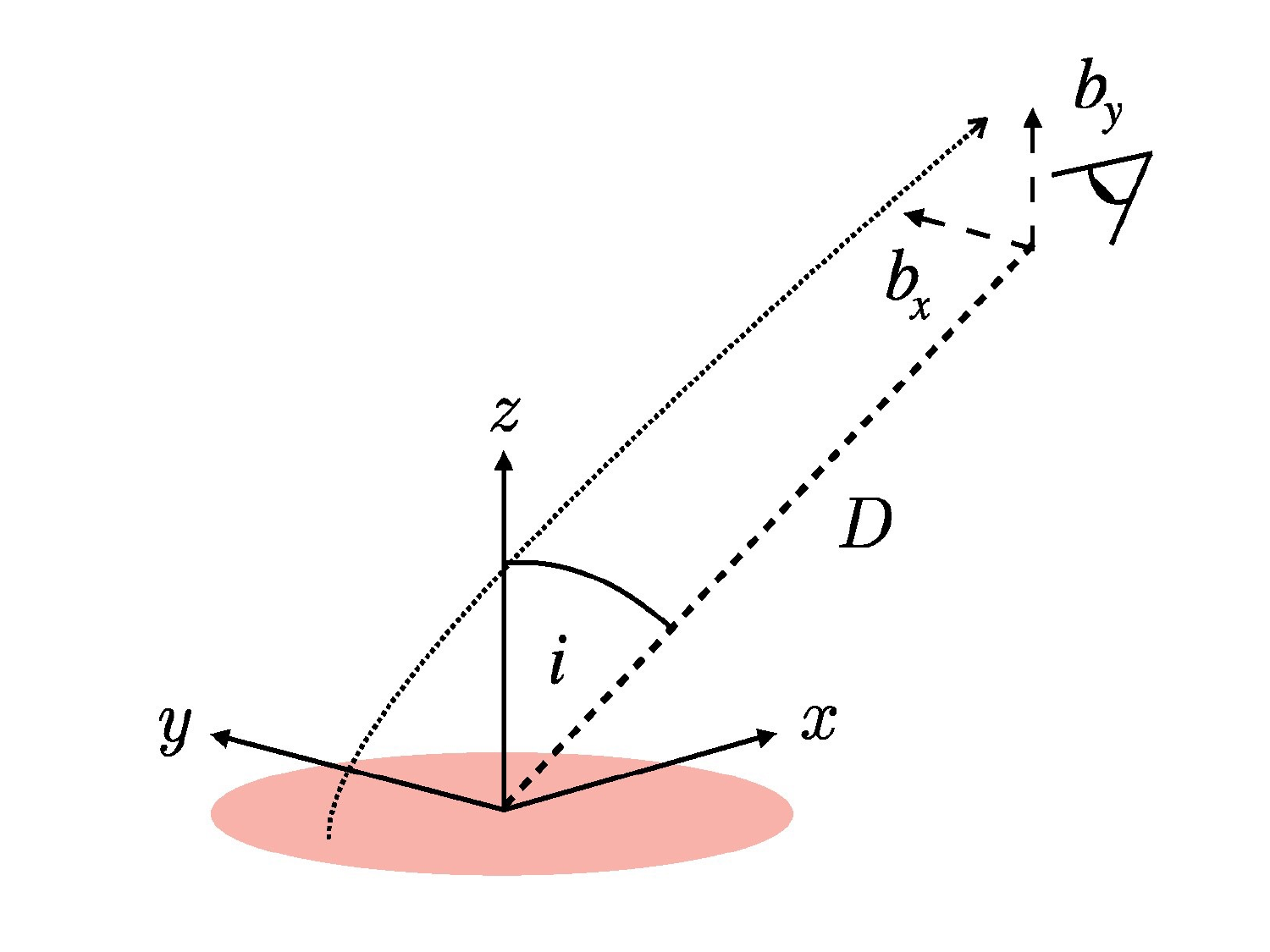}
 
	\caption{Schematics of the ray tracing geometry. The coordinates $b_x$ and $b_y$ lie in the observer plane; $x$ and $y$ in the disk plane. A schematic photon trajectory from the inner disk is shown, observer-disk inclination angle is denoted $i$.}
    \label{fig:schematics}
\end{figure}
Fortunately, analytical Greens function $G_{\Sigma} (r, t; r_0, t_0)$ solutions of this equation are known \citep{Mummery2023b}, provided that the stress tensor $\W$ depends only on radius as a powerlaw 
\begin{equation}\label{eq:stressdef}
    \W = w \left({2r\over r_I}\right)^\mu, 
\end{equation} 
and under the assumption that all disk properties vanish at the ISCO $r_I$ (i.e., a null-stress inner boundary condition). \citet{Mummery2023b}'s analytical solutions describe the  global evolution of an initial ring of total mass $M_{\rm d}$, which is located at $r=r_0$ at an initial time $t = t_0$, and reproduce the full numerical solutions of the relativistic equations with an accuracy at the $\mathcal{O}(1\%)$ level \citep{Mummery2023b}. The functional form of $G_{\Sigma} (r, t; r_0, t_0)$ is also presented in Appendix \ref{app:math} for completeness.

For the stress parametrization used here the definition of the `viscous timescale' is \citep[e.g.,][]{Pringle81}
\begin{equation}\label{eq:t_visc}
    t_{\rm visc} \equiv {2 \over (3 - 2\mu)^2} {\sqrt{GM_\bullet r_0^3} \over  w}\left({r_I \over 2r_0}\right)^\mu.
\end{equation}

Once the black hole parameters and the disk's initial mass and radius are set, the system is fully defined by choosing either \( w \) or the viscous timescale \( t_{\rm visc} \). Typically, \( t_{\rm visc} \) is used as the free parameter due to its clearer physical meaning. In this case, $t_{\mathrm{visc}}$ represents the evolutionary timescale associated with the initial conditions. This does not imply that the disk’s evolutionary timescale remains constant in either time or radius; rather, it will vary with both, as self-consistently described by equation \ref{eq:fund}. Instead, it is $w$ that is assumed to remain constant across time and radius.

While the evolving disk density is not directly observed, its temperature profile—linked to surface density through energy conservation —determines observable properties. The dominant \( r \)-\( \phi \) component of the turbulent stress tensor, \( \W \), governs both outward angular momentum transport (Eq. \ref{eq:fund}) and the conversion of shear energy into heat, which radiates from the surface. In the relativistic case, this local energy dissipation leads to a surface temperature \( T(r, t) \) given by \citep{Balbus2017}:

\begin{figure}[h!]
	\centering
	\includegraphics[width=\columnwidth]{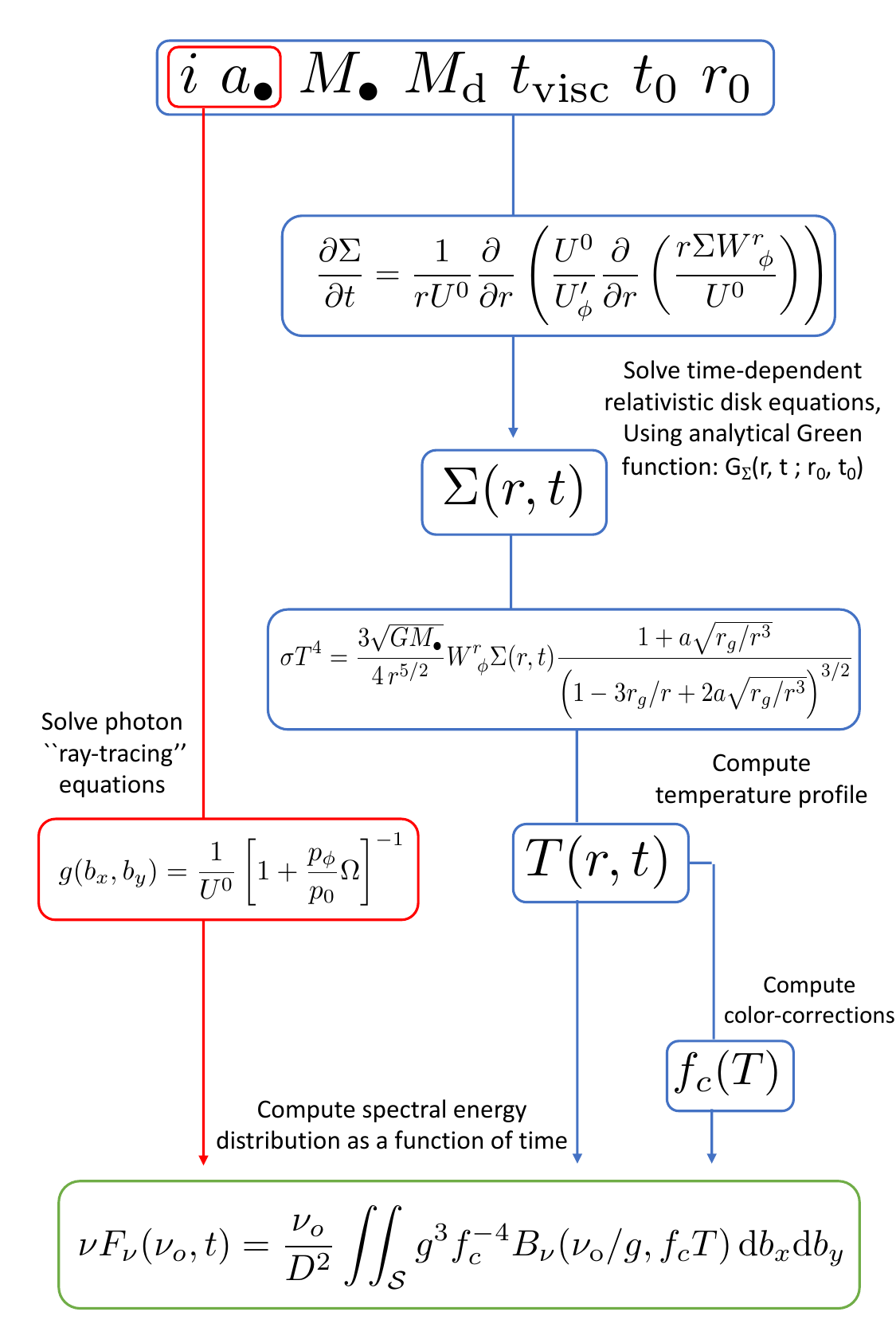}
 
	\caption{Schematics of the implementation of relativist time-dependent disk framework into \texttt{FitTeD}; from the seven free parameters, to the disk's spectral energy distribution as a function of time. See text in \S\ref{sec:theory} for details.}
    \label{fig:fitted_scheme}
\end{figure}
\begin{equation}\label{eq:t}
\sigma T^4 = \frac{3\sqrt{G M_{\bullet}}}{4\,r^{5/2}} \W \Sigma(r, t) \frac{1 + a \sqrt{r_g / r^3}}{\left(1 - {3r_g}/{r} + 2a\sqrt{{r_g}/{r^3}}\right)^{3/2}} ,
\end{equation}
where $\sigma$ is the Stefan-Boltzmann constant. With specific initial conditions, the temperature profile is therefore determined for all radii and future times. 

Assuming an optically thick disk - which should be hold in the range of accretion rates to be explored in this work - the emission is quasi-thermal, with specific intensity in the rest frame:
\begin{equation}\label{eq:int}
    I_\nu (\nu, T) = \frac{1}{f_c^4} B(\nu, f_c T) = \frac{2h\nu^3}{c^2 f_c^4} \left[ \exp\left( \frac{h\nu}{k f_c T} \right) - 1 \right]^{-1} ,
\end{equation}
where $B(\nu, T)$ is the Planck function. In Eq. \ref{eq:int}, $f_c$ is a temperature-dependent color correction factor \citep{Shimura1995,Hubeny2001,Done2012}, owing to the effects of electron scattering and metal opacity in the disk atmosphere. 

The photons emitted with energy $h\nu$ in rest-frame of the disk, are not observed with the same energy, by an distant observer. Instead, by moving through the Kerr spacetime photons both change their energy (as determined by a local observer), and have trajectories which deviate from straight lines (thus changing the observed image plane structure of the source). As a photon field moves through the vacuum, the quantity  $I_\nu/\nu^3$ is conserved \citep{MTW}, meaning that the observed specific intensity is given by

\begin{equation}
    I_\nu^{\rm obs} = \left({\nu_{\rm obs} \over \nu_{\rm disk}}\right)^3 I_\nu^{\rm disk}(\nu_{\rm disk}, T) \equiv g^3 \, I_\nu^{\rm disk}( \nu_{\rm obs}/g, T) . 
\end{equation}
This expression defines the energy shift factor $g$. The energy of a photon, with 4-momentum $p_\mu$, as observed by an observer with 4-velocity $U^\mu$, is given by
    $E = - U^\mu p_\mu$, meaning that the energy shift of the disk photons over their trajectory is given by
\begin{equation}\label{gfac}
    g = {U^\mu_{\rm obs} \, p^{\rm obs}_\mu \over U^\mu_{\rm disk} \,  p^{\rm disk}_\mu} = \frac{1}{U^{0}} \left[ 1+ \frac{p_{\phi}}{p_0} \Omega \right]^{-1} ,
\end{equation}
where $U^0$ and $\Omega$ are evaluated in the disk frame (see Appendix \ref{app:math}), and we have assumed that the distant observer is at rest. The 4-momentum components $p_\phi$ and $p_0$ are conveniently constants of motion in the Kerr metric, and so can be evaluated in the observers frame of reference.

Therefore, the quantity actually observed by the distant observer is:

\begin{equation}
F_{\nu}(\nu_{\rm obs}) = \int g^3 I^{\rm disk}_\nu (\nu_{\rm obs}/g, T) \, \text{d}\Theta_{{{\rm obs} }} .
\end{equation}

\noindent where, $I^{\rm disk}_\nu$ is given by Eq. \ref{eq:int}, and $\Theta_{\rm obs} $ is the differential element of solid angle subtended on the observer’s sky by each element of the disk. In the general form of relativistic case, for an observer at a large distance $D$ from the source the differential solid angle in which the radiation is observed is
\begin{equation}
 \text{d}\Theta_{{{\rm obs} }} = \frac{\text{d}b_x \, \text{d} b_y}{D^2} ,
\end{equation} 
where $b_x$ and $b_y$ are the impact parameters at infinity (i.e., these are Cartesian coordinates which describe the image plane of the telescopes ``camera'').  See schematics in Fig.~\ref{fig:schematics} for further details.

\begin{deluxetable*}{ccCccCCC}
\label{tab:data}
\tablenum{1}
\tabletypesize{\scriptsize}
\tablecaption{Data Summary}
\tablewidth{0pt}
\tablehead{
\colhead{Label} & \colhead{Observatory} &  \colhead{$t_i$} & \colhead{$t_i$} &\colhead{Exp. Time} &\colhead{Total Counts} & \colhead{0.3-2.0 keV Rate$^{(a)}$}& \colhead{ $L_{\rm X}^{(b)}$}  \\
\colhead{} & \colhead{} & \colhead{(MJD)} & \colhead{(Month Year)}& \colhead{} & \colhead{} & \colhead{(Counts s$^{-1}$)}& \colhead{(erg  s$^{-1}$ )} }
\startdata
RO    & ROSAT      &    49533         &   Jun 1994     &               4.8ks             &  \nodata      &       \lesssim 0.005    & \leq 3 \times 10^{40}   \\
Sw1  & \swift/XRT &     55487 ^{+52}_{-37}   &    Sep-Dec 2010       &    8.2ks&        193     &     0.023 \pm 0.002   & 1.10_{-0.09}^{+0.06} \times 10^{43}      \\
X1    & \xmm       &     55532       &    Dec 2010         &   9.8ks&      7332       &       0.745 \pm 0.003   &  1.11_{-0.07}^{+0.08} \times 10^{43} \\
Sw2  & \swift/XRT &      55610^{+37}_{-66}      &  Jan-Mac 2011           &    13.6ks&     278        &     0.020 \pm 0.001     & 9.89_{-0.71}^{+0.75} \times 10^{42}    \\
Sw3  & \swift/XRT &      56560^{+2}_{-4}      &    Sep 2013        &     5.9ks&    78         &     0.013 \pm 0.001     &8.32_{-0.32}^{+0.46} \times 10^{42} \\
X2$^{(d)}$    & \xmm       &     56996        &    Dec 2014       &        65.5ks&        28605     &     0.437 \pm 0.003    &7.08_{-0.40}^{+0.41} \times 10^{42} \\
X3$^{(d)}$     & \xmm       &      58476       &   Dec 2018          &      28.4ks&     6231        &     0.219 \pm  0.003    & 3.96_{-0.21}^{+0.29} \times 10^{42} \\
X4    & \xmm       &    58633         &     Jun 2019        &     69.5ks&      8822       &     0.157 \pm  0.003   &3.65_{-0.11}^{+0.15} \times 10^{42} \\
\enddata
\tablecomments{(a) This is an instrument dependent quantity, should only be compare between observations of the same instrument. (b) Measured in 0.2-2.0 keV range. As every X-ray flux/luminosity these are model-dependent quantities. (d) Contemporaneous \hst/UVIS UV spectra available.}
\end{deluxetable*}

Note that $g$ depends upon $b_x$ and $b_y$ (through the 4-momentum of the observed photon, eq. \ref{gfac}).  There is no simple analytical solution which relates the quantity $g$ and the photon's emitted radius $r$ to the quantities $b_x$ and $b_y$ for general Kerr spin parameter $a_{\bullet}$ and observer inclination $i$. This calculation must therefore be performed numerically. The numerical `ray-tracing' algorithm presented and detailed in sections 2.3.2 of \citet{Mummery2024fitted} is used, and is refereed to for details.

Finally, combination all these effects, the flux observed from disk surface $\cal S$ is written as:

\begin{equation}\label{eq:flux}
\nu F_\nu (\nu, t) = \frac{\nu _{\rm obs}}{D^2}   \iint_{\cal S} {g^3 f_{c}^{-4} B (\nu_{\rm obs} / g , f_{c} T)}\,  {\text{d}b_x  \text{d} b_y},
\end{equation}

\noindent where we express the dependence of $\nu F_\nu$ on both time and frequency to highlight the capabilities of \texttt{FitTeD} implementation, which we now aim to extend to X-ray spectroscopy. 

Given seven free parameters:
$M_{\bullet}$ (black hole mass),
$a_{\bullet}$ (dimensionless black hole spin),
$M_d$ (initial disk mass)
$t_{\rm visc}$: (`viscous' time-scale),
$t_0$ (disk formation time),
$r_0$ (initial disk radius),
$i$ (observer-disk inclination angle), that describe the system, the resulting spectrum is fully determined across all frequencies and times. A schematic summary the process from the free parameters, through all computations, to the final spectral solution is illustrated in Fig.~\ref{fig:fitted_scheme}. This provides a tractable model that can be fitted to multi-wavelength data—whether photometry or spectroscopy—across different energy ranges, such as UV/optical or X-rays, and at multiple epochs, in a fully self-consistent manner.

Of course, this framework assumes a smooth disk evolution, valid on timescales not much shorter than the disk’s viscous timescale. While this is a ubiquitous feature of all analytical modeling of accretion disks, real accretion flows are turbulent, violating this assumption on shorter timescales. As a result, short-timescale fluctuations in the disk light curves are not captured \citep[e.g.,][]{Pasham2024,Yao2024_22lri}, leading to potentially large discrepancies, especially in TDE X-ray light curves where temperature fluctuations from turbulence are exponentially amplified into large luminosity fluctuations in the Wien-tail \citep{MummeryTurner2024}. Nevertheless the long-term evolution of the accretion flow, which is the subject of this paper, should be fully captured by the model (and if it is not, this is also an important result). 

For a specific implementation of this framework, the radial dependence of the stress tensor, \( \mu \), in Eq. \ref{eq:stressdef}, and the color-correction factor (\( f_c \)) must be chosen, and any specific choice will naturally be somewhat \textit{ad hoc}. Following the original \texttt{FitTeD} implementation, we assume \( \mu = 0 \), a choice that has been numerically verified to be subdominant compared to statistical uncertainties in the data \citep{Mummery2024fitted}. Regarding \( f_c \), while the original \texttt{FitTeD} adopts the prescription of \citet{Done2012}, here we use the analytical expression from \citet{Chiang2002}, calibrated on the numerical simulations of \citet{Hubeny2001}. This choice ensures consistency with the results of \citetalias{Guolo2025} on \target.   Any potential impact of these choices on our conclusions regarding \target\ will be discussed in \S\ref{sec:limitations}.

It is also important to define some quantities that are neither free parameters nor directly observed but can be inferred from a full disk solution, and which will be useful later. The bolometric luminosity of the disk can be simply computed from Eq. \ref{eq:flux} as
\begin{equation}
     L_{\rm Bol}(t) 
     = 4\pi  \int_{\rm disk} r \, \sigma T^4(r, t) \, {\rm d}r, 
\end{equation}
which can also be expressed in terms of the Eddington luminosity ($L_{\rm Edd}$), defining the Eddington ratio as  
\begin{equation}
    \lambda_{\rm Edd} = \frac{L_{\rm Bol}}{L_{\rm Edd}}.
\end{equation}

Another key quantity is the mass accretion rate, which, for a time-dependent disk (i.e., not in steady state), depends on both time and radius:  
\begin{equation}
     \dot{M}(r, t) \equiv 2 \pi r U^r \Sigma 
    = - \frac{2\pi  U^0}{U_\phi'} \frac{\partial}{\partial r} \left(\frac{r \Sigma \W}{U^0}\right) .
\end{equation}
       
\noindent The accretion radiative efficiency ($\eta$) is equal to
\begin{equation}
    \eta(a_\bullet) = 1 - \left(1 - \frac{2r_g}{3r_I}\right)^{1/2}.
\end{equation}

\noindent However, in a time-dependent disk system, the bolometric luminosity is not simply related to the accretion rate via the standard efficiency factor, i.e.,  $L_{\rm Bol} \neq \eta \dot{M}(r_I) c^2$. Nevertheless, these two quantities asymptotically approach each other at late times as the system evolves toward a quasi-steady state.

The reader may note there is a large amount of overlap between this section and the Section 2 of the original \texttt{FitTeD} paper \citep{Mummery2024fitted}, where further details on the code’s underlying computations can be found. For a more comprehensive discussion of the physics and derivations of relativistic time-dependent disk theory, we refer to \citet{Eardley1975,Balbus1999,Balbus2017,Mummery2020,Mummery2021,Mummery2024}. We now turn to \target, the dataset, and our approach to implementing \texttt{FitTeD} into \texttt{pyXspec}.

\section{Data and Fitting Setup}\label{sec:data}

\subsection{Data}
Observations at the position of \target were conducted in the 1990s by \rosat, revealing no detections. The latest and deepest observation in June 1994 (MJD 49533, hereafter $t_{\rm RO}$) with a $\sim$5 ks exposure securely excluded the presence of any source with a 0.2–2.0 keV flux (luminosity) higher than $\sim 5\times10^{-14}$ ($\sim {\rm few} \times 10^{40}$ \ergs). In July 2010 (MJD , hereafter $t_{\rm X0}$), a $\sim$6-second observation of the field by the \xmm\ slew survey \citep{Saxton2011} detected a new source with a 0.2–2.0 keV count rate of $1.4 \pm 0.4$ counts s$^{-1}$, corresponding to a 0.2-2.0 keV flux (luminosity) of $\sim 1\times10^{-12}$ \fluxunit ($\sim 1.5\pm0.5 \times 10^{43}$ \ergs).

In the following years, numerous follow-up observations were conducted using both the X-ray Telescope (XRT) aboard \swift\ and long-exposure observations with \xmm. In this work, we focus on data collected by these missions from discovery until early 2020, during which the source exhibited a smooth, declining evolution \citep{Miniutti2023_longterm}.

The X-ray data used in this work is summarized in Table \ref{tab:data}, and shows the naming convection we will adopt to refer to each observation. Throughout these observations, the source has exhibited a thermal-like, extremely soft spectrum, with all detected photons having energies below 2.0 keV. The reduction and processing of the X-ray data are detailed in Appendix \ref{app:data}, which primarily consists of: i) Stacking multiple \swift\ snapshots to create high signal-to-noise (S/N) spectra, thereby averaging out short-term variability that cannot be captured by the analytical model described in \S\ref{sec:theory}. ii) Removing QPE emission present in X3 and X4 (see small panel in Fig.~\ref{fig:lc+SED}) to isolate the long-term `quiescent' disk emission for modeling. The QPEs were not present in X2 (see small panel in Fig.~\ref{fig:lc+SED}), while X1 exposure time was not long enough to constraint the presence of the eruption.


Ultraviolet (UV) spectra (1150–3000 \AA) from the \textit{Hubble Space Telescope} (\hst) are available at two epochs: December 2014 and December 2018, therefore contemporaneous to X2 and X3 in Table \ref{tab:data}. These spectra were recently analyzed by \citetalias{Guolo2025}, and it is useful to summarize some key findings: i) The UV spectra are predominantly continuum-dominated (i.e., ‘featureless’), with an intrinsic spectral shape given by $\nu F_{\nu} \propto \nu^{4/3}$ to first order; ii) There was a $\sim 9\pm2 \%$ decrease in far-UV flux between the two epochs; iii) The diffuse stellar contribution to the spectra is subdominant ($\leq 5\%$ of the UV flux at the observed epochs); iv) The spectral energy distribution (SED), combining X-ray and UV spectra, can be well described by a thin accretion disk model; v) the disk was as extended as $R_{\rm out}$ of $\mathcal{O}(10^3\,r_{\rm g})$, with a $\sim 10\%$ expansion from 2014 to 2018.

In this work, we use the same UV spectra data as reduced and described in \citetalias{Guolo2025}. To avoid unnecessary recomputation in the overall modeling, we subtract the underlying stellar population as already modeled in \citetalias{Guolo2025}. Additionally, we bin the spectra using synthetic narrow-band filters, as detailed in Appendix \ref{app:data}. This process preserves all relevant information, as the spectra are continuum-dominated, and full wavelength coverage is maintained, while reducing substantially the number of energy bins the model will need to compute.

\subsection{\texttt{FitTeD-XSPEC}}\label{sec:fitted_xspec}
Integrating a physical or phenomenological model into an X-ray spectral fitting package like \texttt{XSPEC} (or \texttt{pyXspec}) requires the flux per energy bin, $F_{E}(E)$ or $F_{\nu}(\nu)$, to be expressible as an analytical or semi-analytical function of the model’s free parameters. This allows convolution with the instrument response for a direct comparison of predicted and observed photon counts, enabling parameter optimization via minimization or Bayesian inference.

In standard fitting, spectra are computed independently for each epoch, with free parameters for each \texttt{DataGroup} (e.g., \texttt{pyXspec} data sets). For example, when fitting a blackbody function, each \texttt{DataGroup} has two free parameters $T_{\rm BB}$ and $R_{\rm BB}$. Constraints can be imposed across epochs, such as tying $R_{\rm BB}$ but allowing $T_{\rm BB}$ to vary, if that is physically justifiable. Nevertheless, each \texttt{DataGroup} still retains independent parameters, which can vary freely between epochs, with no physically-driven basis for such time evolution.

As shown in \S\ref{sec:theory} and Fig.~\ref{fig:fitted_scheme}, our framework generates a computable spectrum based on free parameters. The resulting spectrum depends not only on energy but also on time, producing an evolving spectrum at each time realization, $t_i$. The \texttt{FitTeD} implementation in \texttt{pyXspec} uses seven fundamental free parameters—$M_{\bullet}$, $a_{\bullet}$, $i$, $M_{\rm d}$, $t_{\rm visc}$, $t_0$, and $r_0$—shared across epochs. An additional parameter, $t_i$, determines the time at which the evolving spectrum is computed; however, it is not a free parameter but is instead fixed based on the observational timestamps of the dataset (e.g., as given in Table \ref{tab:data}). While the distance $D$ is in essence a free parameter of the model, it is here fixed by the redshift and the assumed cosmology.
This ensures that the observed spectra at different $t_i$ can be directly compared to the model-predicted spectra, $F_{\nu}(\nu, t_i)$ (Fig.~\ref{fig:fitted_scheme}). In this sense, the model is fitting both the spectra and the light curve of the system simultaneously. The optimal combination of the seven parameters that best describes the data across all epochs and energy bins can then be determined using parameter inference techniques such as Nested Sampling or Markov Chain Monte Carlo.\footnote{Given the high dimensionality of the problem and the large dataset required to constrain the parameters, the standard Levenberg–Marquardt minimizer available in \texttt{XSPEC} is unlikely to achieve global convergence \citep{Buchner2023}.} The model extends to lower wavelengths as well. Eq. \ref{eq:int} enables flux computation at any energy, and \texttt{pyXspec} allows a \texttt{DataGroup} to include multiple spectra for simultaneous fitting of UV and X-ray data, as available for \target's in 2014 and 2018.

We wish to stress that while this approach adds additional free parameters to the model of the first X-ray spectrum taken from a source (i.e., one moves from $M_{\bullet}$, $a_{\bullet}$, $i$ and $\dot M$ in a classical relativistic steady state model, e.g.,  \texttt{kerrBB} \citet{Li2005}, to $M_{\bullet}$, $a_{\bullet}$, $i$, $M_{\rm d}$, $t_{\rm visc}$, $t_0$, and $r_0$ here), one gains both a much more robust physical model of the disk (a disk described by a constant $\dot M$  by definition can never change with time, as it is a steady state solution) but more importantly {\it reduces} the number of free parameters in the fitting procedure of a source with {\it multi-epoch} data. Classical sequence-of-steady-state relativistic data fitting approaches are described by $N_{\rm par}=3+N_{\rm obs}$ free parameters for $N_{\rm obs}$ observations (i.e., the black hole mass, spin and observer inclination do not (or should not) change, but $\dot M$ is assumed to be different in every epoch). Our approach uses $N_{\rm par}=7$ free parameters {\it independent} of the number of epochs studied. For sources with many epochs of data, this is a vast improvement in model complexity, which should always be minimized when fitting models to data. For our work, GSN 069 has $N_{\rm obs} = 7$ X-ray spectra, reducing the total number of free parameters of a relativistic model from $10$ for a ``sequence of steady states'', to $7$. 

The improvement is even more significant when compared to the standard approach commonly used in TDE literature for X-ray spectral modeling, such as the use of \texttt{diskbb} \citep{Mitsuda1984, Makishima_86}. Aside from being outdated relative to current theoretical and computational advances (as e.g., it has neither relativist nor radiative transfer corrections), \texttt{diskbb} is often incorrectly assumed to represent a standard \citet{Shakura1973} disk—though it does not, as shown in detail by \citet{Zimmerman2005}. 
Crucially for our purposes, \texttt{diskbb}-based fits are typically performed independently at each epoch, with both parameters allowed to vary freely, resulting in $N_{\rm par} = 2 \times N_{\rm obs}$. While we do expect the ‘temperature’ parameter to vary between epochs (in fact, on even shorter timescales than allowed in \S\ref{sec:theory}; see, e.g., \citealt{MummeryTurner2024}), there is no physical justification for the normalization (which depends only on the inner disk radius and inclination) to vary freely between epochs, if the X-rays are thought to originate from direct emission of an stable corona-less accretion disk.

\subsection{Fitting setup for \target}
\begin{figure*}[]
	\centering
	\includegraphics[width=0.8\textwidth]{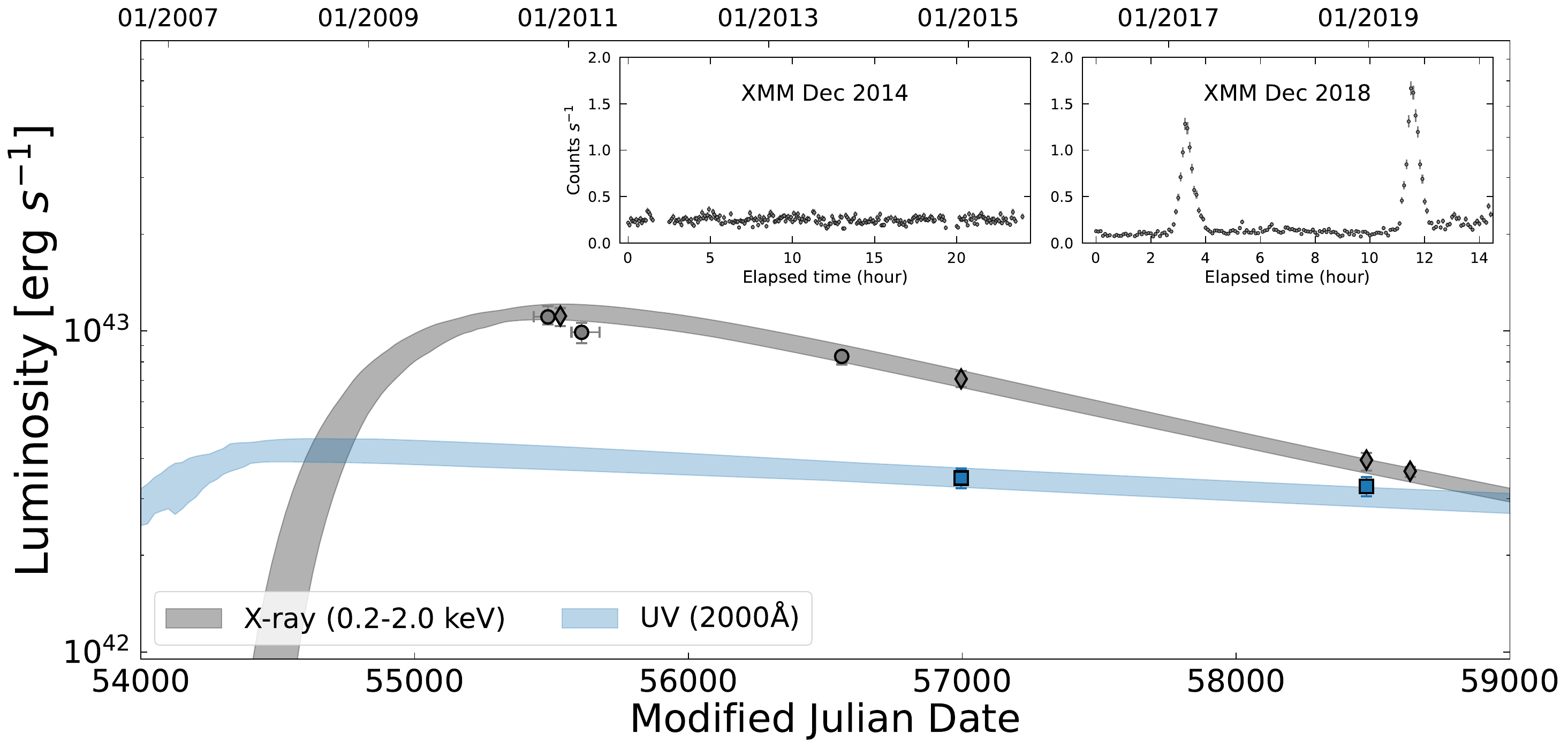}\\
    \vspace{1cm}
    \includegraphics[width=1\textwidth]{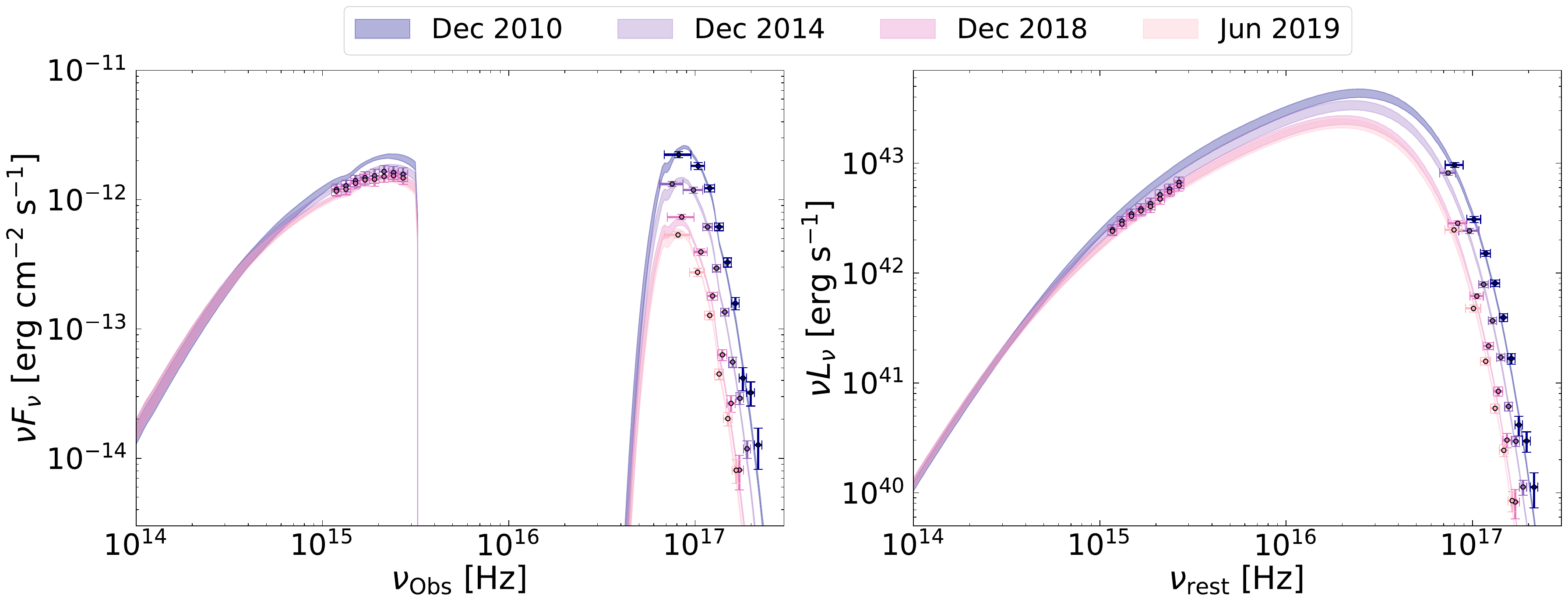}
	\caption{Results of \texttt{FitTeD-XSPEC} fitting to \target data, including seven X-ray spectra and two UV spectra from 2010 to late 2019.  \textbf{Top:} Evolution of the 0.3--2.0\,keV X-ray luminosities (gray) and UV luminosities at 2000\,\AA\ (blue). Shaded regions represent the model-predicted values, at the 68\% credible intervals, while the data points show realizations at the observed epochs. Gray symbols indicate data from \xmm\ (diamonds), \swift/XRT (circles), and the \xmm\ Slew Survey (star). Note that the Slew Survey data was not included in the fit, but is nonetheless reproduced by the model within uncertainties.  
    \textbf{Bottom:} Spectral energy distributions at the epochs of the \xmm\ observations (and \hst, in 2014 and 2018). The left panel shows observed fluxes (including absorption and extinction), while the right panel shows intrinsic luminosities corrected for both. In both panels the X-ray spectra are unfolded to the maximum a posteriori solution.
 }
    \label{fig:lc+SED}
\end{figure*}

\begin{figure*}[]
	\centering
	\includegraphics[width=0.9\textwidth]{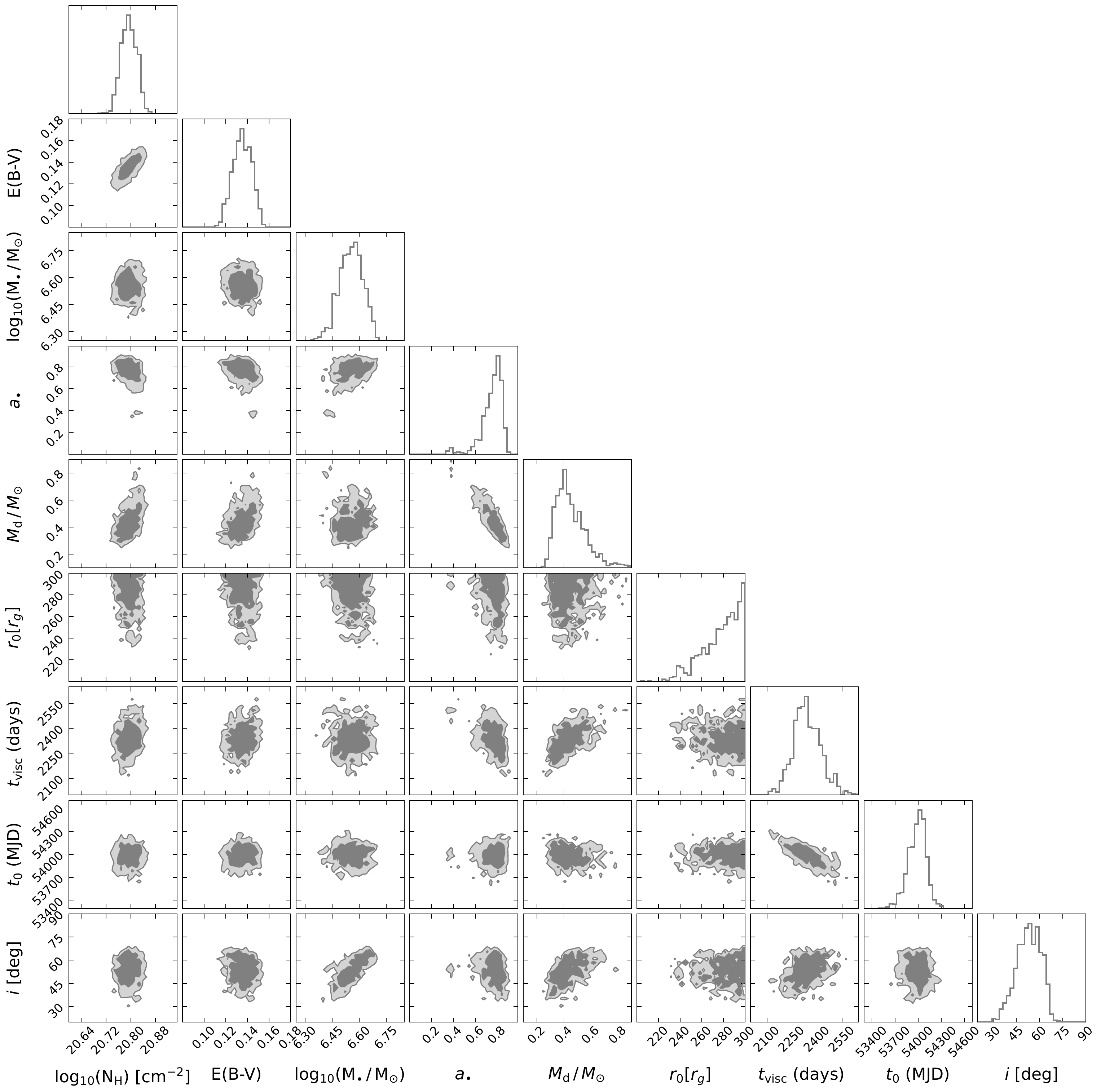}\\
    \caption{Corner plot of the marginalized posterior from \texttt{FitTeD-XSPEC} fit of \target. The contours are 68\% and 99\% credible regions.}
    \label{fig:corner}
\end{figure*}

\begin{figure*}[]
	\centering
	\includegraphics[width=0.9\textwidth]{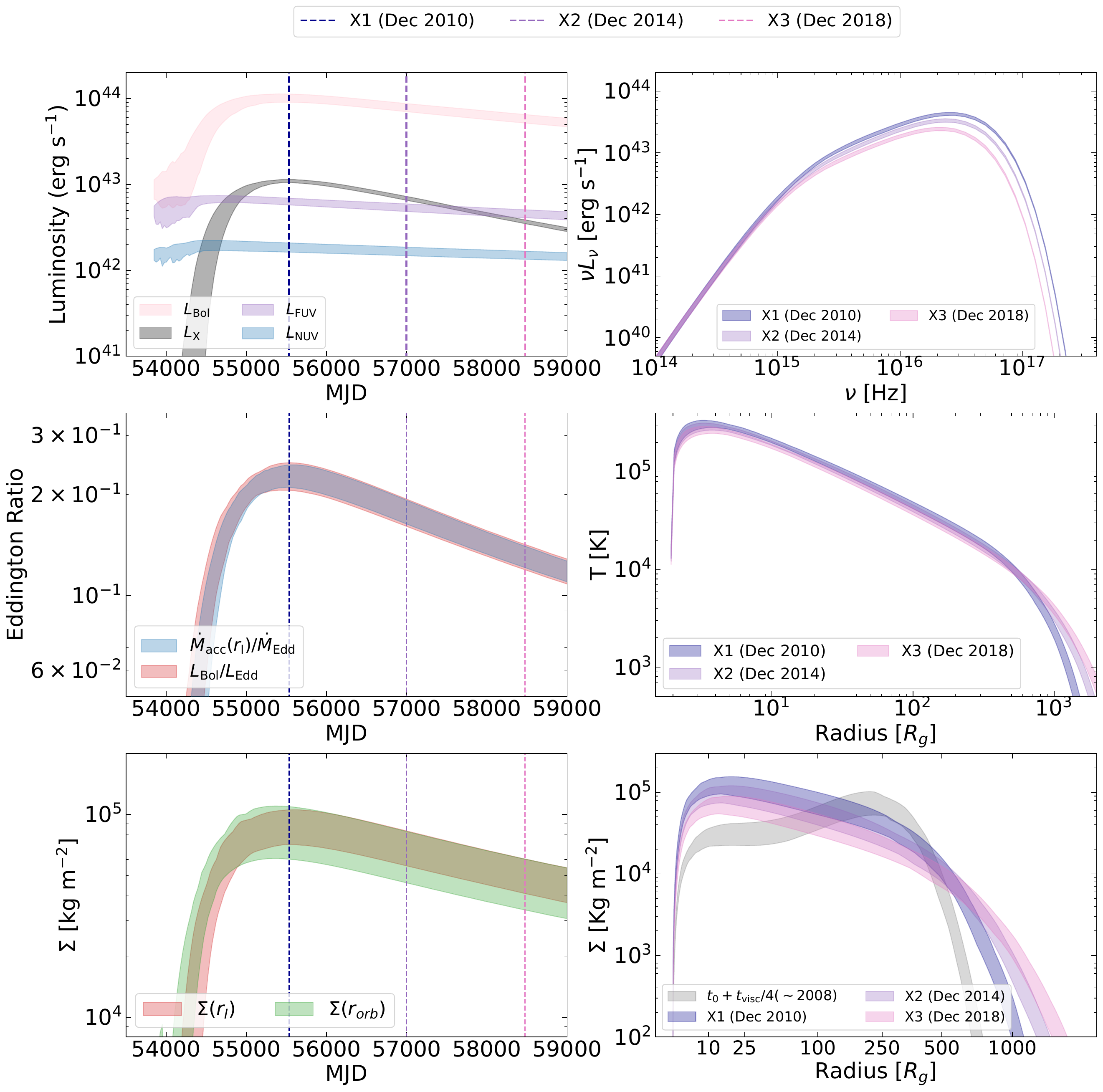}\\
    \caption{Evolution of the physical properties of \target's accretion disk. \textbf{Left:} Time evolution of key disk properties, from top to bottom: monochromatic luminosities, Eddington ratios, and surface density (at ISCO and at $r_{\rm orb}$, see \S\ref{sec:qpes}). \textbf{Right:} Disk properties at representative epochs: spectral energy distribution (top), radial temperature profile (middle), and radial surface density profile (bottom). Different colors correspond to distinct epochs as defined in Table~\ref{tab:data}.} 
    \label{fig:profiles}
\end{figure*}

The data fitting begins by loading the seven X-ray spectra (X1–X4 and Sw1–Sw3) listed in Table \ref{tab:data}, each assigned to a separate \texttt{DataGroup} with its corresponding observation timestamp. For X2 and X3 (Dec 2014 and Dec 2018), we also incorporate the contemporaneous UV spectra using HEASoft’s ``ftflx2xsp" tool (v6.33.2; \citealt{Heasarc2014}), which generates the necessary diagonal response files.

Since \texttt{XSPEC} does not support Poisson statistics for UV/optical data, we rebin the X-ray spectra before loading them. This is done using an `optimal binning' scheme \citep{Kaastra2016}, ensuring each bin contains at least 10 counts (\texttt{grouptype=optmin}, \texttt{groupscale=10} in \texttt{ftgrouppha}). Given the high total counts per spectrum (see Table \ref{tab:data}), each bin typically contains \(\gg\) 10 counts, allowing the simultaneous X-ray and UV fit to be performed using Gaussian statistics. A conservative 5\% systematic uncertainty is added to all datasets, to a account for possible miss calibration between missions and to help account for short-term variability that cannot be physically reproduced by the model.

To model dust attenuation in the host galaxy, we use the \texttt{reddenSF} model \citep{Guolo_Mummery2025}, based on the \cite{Calzetti2000} attenuation law, with color excess \( E(B-V) \) as the free parameter. Galactic extinction is modeled separately using \texttt{XSPEC}’s native \texttt{redden}, which follows \cite{Cardelli1989}. Neutral gas absorption is handled with two \texttt{phabs} components: one fixed to the Galactic value and another redshifted to the host frame. The intrinsic model is shifted to the source rest frame using \texttt{zashift} with \( z = 0.018 \).

In \texttt{XSPEC} notation, the full model is: \texttt{phabs$\times$redden$\times$zashift(phabs$\times$reddenSF$\times$FitTeD)},
where Galactic absorption is fixed at \( N_{\rm H, G} = 2.3 \times 10^{20} \) cm\(^{-2}\) \citep{HI4PI2016}, and Galactic extinction at \( E(B-V)_{\rm G} = 0.023 \) \citep{Schlafly2011}. The final model, to simultaneously describe all seven X-ray spectra and the two UV spectra, has therefore nine free parameters: seven from the time-dependent disk model, plus the intrinsic color excess \( E(B-V) \) and intrinsic hydrogen column density \( N_{\rm H} \).

Parameter inference is performed using Bayesian X-ray Analysis (BXA) v4.0.7 \citep{Buchner2014}, which integrates the \texttt{UltraNest} nested sampling algorithm \citep{Buchner2019} with \texttt{PyXspec}. We assume linear or log-linear priors for all of the  parameters, except for \( E(B-V) \), in which a Gaussian prior centered at \( 0.10 \) with a standard deviation of \( 0.02 \) was assumed. This is motivated by independent constraints from the Balmer decrement in the host galaxy’s optical spectrum \citep{Wevers2024_qpehost}, which resulted in E(B-V)$ = 0.10\pm0.02$. However, instead of fixing its value, we use a Gaussian prior to propagate its uncertainties into the posterior distributions of other parameters, and to allow for some small deviations from the Balmer decrement value.

A crucial factor is the selection of boundaries for the flat priors of parameters like \( r_0 \) and \( t_0 \), guided by both observations and physical expectations. As discussed in Appendix \ref{app:data}, the X-ray flux (or effective area-normalized count rate) remains consistent across X0, Sw1, and X1 within uncertainties. While this contrasts with \citet{Miniutti2023_longterm}, it is consistent with \citet{Shu2018}. This implies that \target’s X-ray rise and peak emission may have been missed, which can limit our ability to constrain key system parameters.

Nonetheless, we do not impose a fixed peak date in our modeling. Instead, we provide the data and set a prior bound on \( t_0 \) based on available constraints: since \rosat\ did not detect the source in 1994 and the source was discovered with the \xmm Slew observation in 2010, we have that  
$t_{\rm RO} \leq t_0 \leq t_{\rm X0}$, with a flat prior in this range.

Given the TDE nature of \target \citepalias[see][for a detailed discussion]{Guolo2025}, the initial disk radius ($r_0$) is expected to form very compactly. In a simplistic picture, it should form near the so-called `circularization radius' ($R_{\rm circ}$), which is approximately twice the periapsis radius of the disrupted star, assuming a full disruption. For a main-sequence star of mass $M_{*}$, this radius can be expressed as:

\begin{equation} 
R_{\rm circ} \approx 92 \, \beta^{-1} \left(\frac{M_*}{M_{\odot}} \right )^{7/15}\left( \frac{M_{\bullet}}{10^6 M_{\odot}} \right )^{-2/3} r_g
\end{equation}

\noindent where $\beta$ is the impact parameter. Rewriting this in terms of \texttt{FitTeD} parameters gives:

\begin{equation}
    R_{\rm circ} \approx \frac{92}{ \beta f_d^{7/15}} \left(\frac{M_d}{M_{\odot}} \right )^{7/15}\left( \frac{M_{\bullet}}{10^6 M_{\odot}} \right )^{-2/3} r_g
\end{equation}

\noindent where the disk mass $M_{\rm d}$ is written as a fraction $f_d$ of $M_*$, and $f_d \leq 1/2$ \citep{Rees1988}. While the probability distribution of $\beta$ for a full disruption peaks at $\beta = 1$ \citep{Stone2016}, in principle $\beta$ can be arbitrarily large, and $f_d$ arbitrarily small—owing to the complex and potentially inefficient dynamical process that governs how the returning mass settles into a disk. Consequently, $R_{\rm circ}$ can, in principle, for combinations $\beta$ and $f_d$, take any of value higher than $r_I$. Since only $M_{\rm d}$ and $M_{\bullet}$ are free parameters in our disk model, $r_0$ cannot therefore be uniquely determined from only these two parameters and must instead be treated as an independent, free parameter with an arbitrarily chosen bound. 

The earliest observational constraint on the disk's physical size comes from the 2014 SED fitting \citepalias{Guolo2025}, which found an somewhat extended outer radius, $R_{\rm out}$ of $\mathcal{O}(10^3\,r_{\rm g})$ \footnote{This 2014 disk size is an entirely observationally driven result \citepalias{Guolo2025}, and results from the fact that the GSN 069 UV spectra in 2014 was already dominated by the ``mid frequency'', i.e. $\nu L_{\nu} \propto \nu^{4/3}$, range of a standard disk spectrum, instead of the Rayleigh–Jeans tail $\nu L_{\nu} \propto \nu^{3}$. This means that the outer disk edge was larger than the typical radial scale of disk material which emits UV photons, providing a lower bound on the disk size.}. It is therefore reasonable to expect that $r_0$ should be smaller than this value, from which the disk will expand over time owing to angular momentum conservation (Eq. \ref{eq:fund}). We therefore assume a linear prior on $r_0$ with the range $r_I \leq r_0 \leq  300 \, r_g$.

We wish to stress that when one does not have good data constraining the rise to X-ray peak of a disk source, one cannot (and should not) interpret the parameters $r_0$ and $t_0$ literally. The initial condition used in this model is naturally idealized, something which is required for efficient computation of disk spectra. As the governing equation is diffusive in character, information about the chosen initial condition is quickly lost, and therefore with only post-peak data it is  unlikely that $r_0$ and $t_0$ will correspond to truly physical values. Some freedom in the fitting range of $r_0$ and $t_0$ should be permitted, so as to not over-constrain the first few spectral fits, but too much freedom will lead to poor convergence. We found an upper bound of $\sim 300 r_g$ to be a good compromise between these two modeling effects. We verified that our choices of priors on $r_0$ and $t_0$ did not change the posterior distributions of the other, more physical, parameters by meaningful amounts. 

We now discuss the  results of the fitting of our \texttt{pyXspec} \texttt{FitTeD} implementation to the seven X-ray and two UV spectra of \target.

\section{Results and Discussion}\label{sec:discus}
\subsection{Results}\label{sec:results}

The results of our fitting to \target are presented in Figures \ref{fig:lc+SED}, \ref{fig:corner}, and \ref{fig:profiles}. The top panel of Figure \ref{fig:lc+SED} displays the X-ray (0.3–2.0 keV) light curve, including both the measured values at the times of the seven spectra (obtained using \texttt{XSPEC}'s \texttt{calcLumin}) and the model prediction over time. This panel also includes the UV light curve at 2000\AA, derived from the two \textit{HST} spectra, alongside the model prediction. Additionally, inset panels show the intra-observation X-ray light curves for X2 (2014) and X3 (2018), highlighting the emergence of QPEs in 2018. 

The bottom panels provide a full view of the spectral energy distribution (SED) evolution from 2010-2019 (X1-X4). The left panel shows the `observed' (without corrections for Galactic or intrinsic absorption/extinction) SED, and includes the UV spectra from the two available epochs and three unfolded \textit{XMM-Newton} spectra. In contrast, the right panel displays the intrinsic SED, fully corrected for absorption and extinction effects. Fig. \ref{fig:corner} shows the corner plot of the marginalized posterior of the free parameters of the modeling.

In Fig.~\ref{fig:profiles}, we present the time evolution of various model properties and observables. The left panels display the evolution of monochromatic ($L_{\rm X}$, $L_{\rm FUV}$, $L_{\rm NUV}$) and bolometric ($L_{\rm Bol}$) luminosities (top left), Eddington ratios ($\dot{M}_{\rm acc}/\dot{M}_{\rm Edd}$ and $L_{\rm Bol}/L_{\rm Edd}$, middle left), and disk surface density, $\Sigma(r_i, t)$ (bottom left), at two characteristic radii of interest (see, e.g., \S\ref{sec:qpes}). The right panels show key properties at selected characteristic times of the \target’s evolution. From top to bottom, we show the radial temperature profiles, $T(r, t_i)$, the spectral energy distributions, $\nu L_{\nu} (\nu, t_i)$, and the mass surface density profiles, $\Sigma(r, t_i)$. We now discuss the parameters inferred from our modeling.

First, we consider the parameters that are common between this time-dependent approach and the time-independent modeling performed in \citetalias{Guolo2025}. The intrinsic host galaxy parameters we recover, ${\rm log} (N_{\rm H}/{\rm cm^{-2}}) = 20.80 \pm 0.05$ and $E(B-V) = 0.13\pm0.01$, are slightly higher -- expected given the slightly distinct properties of the dataset and the nature of the model -- but consistent at $\leq 2\sigma$ level, with the results of \citetalias{Guolo2025}. Also relevant to notice, that while we have not tied in any form  $E(B-V)$ and $N_{\rm H}$, their best fitted values, are consistent to the first order, to the known Galactic gas-to-dust ratio \citep{Guver2009}.

Regarding the black hole spin parameter ($a_{\bullet}$), \citetalias{Guolo2025} found a bimodal posterior distribution when fitting the 2014 and 2018 SEDs, with one mode favoring low or slightly negative $a_{\bullet}$ and another favoring high $a_{\bullet}$, though no region of parameter space was definitively excluded. Our time-dependent modeling yields $0.6 \leq a_{\bullet} \leq 0.9$, aligning with the high-spin mode of \citetalias{Guolo2025}.  

A natural question is why the low-spin mode from \citetalias{Guolo2025} is now excluded.
We believe that the following two key differences likely broke the previous modeling degeneracy: (i) our study incorporates seven X-ray spectra, compared to only two in \citetalias{Guolo2025}, and (ii) in a time-dependent disk model, $a_{\bullet}$ affects not only the observed spectrum (Eq.~\ref{eq:flux}), as in the time-independent case, but also the disk’s fluid evolution via the surface density evolution equations (Eq.~\ref{eq:fund}). Together, these factors 
resulted in a uni-modal distribution, favoring high—but not maximal—spin. In \S\ref{sec:limitations}, we discuss why these $a_{\bullet}$ values should always be interpreted with caution.

\begin{figure}[h!]
	\centering
	\includegraphics[width=\columnwidth]{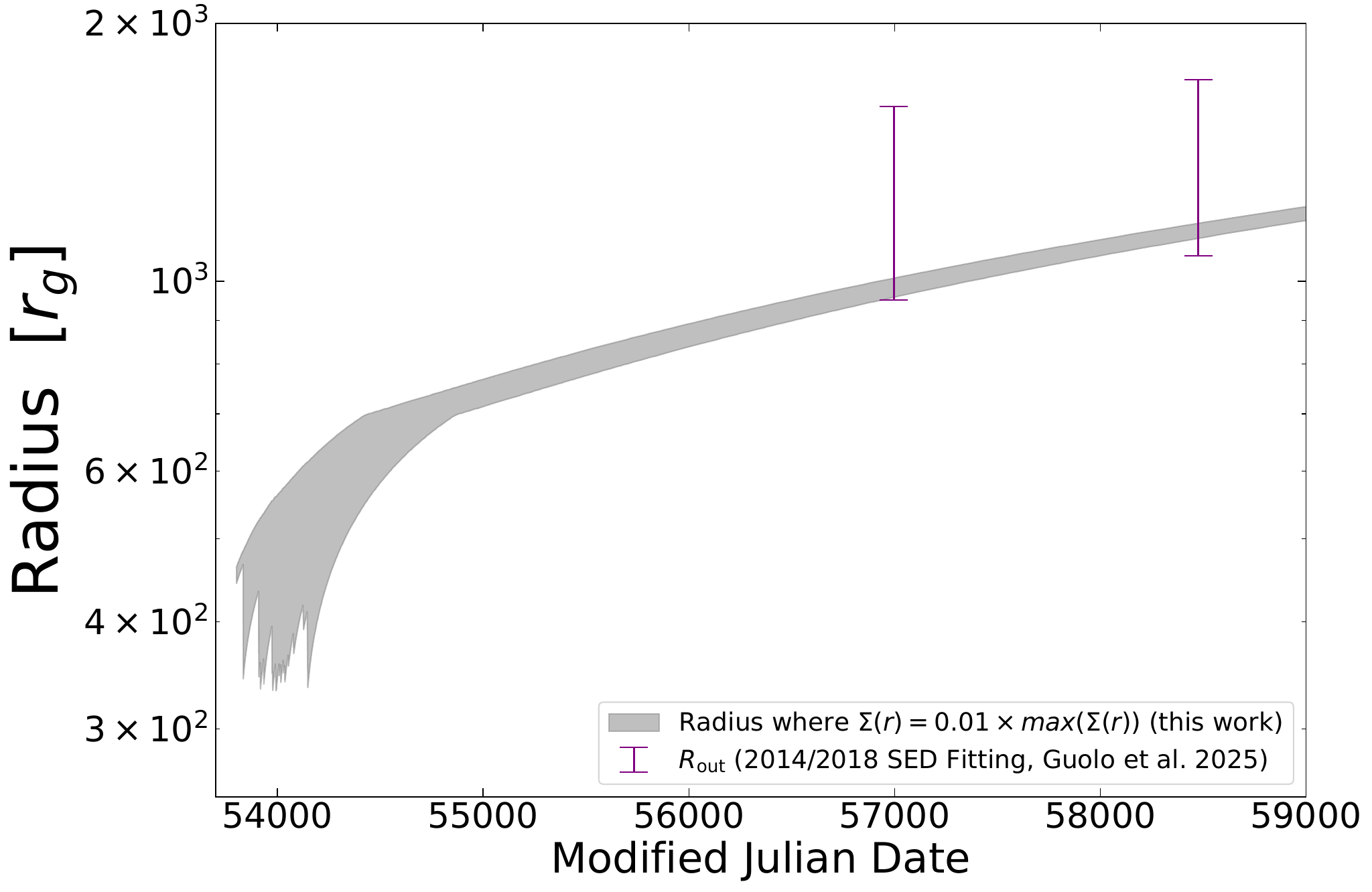}
 
	\caption{Comparison between results of the time-dependent modeling (\texttt{FitTeD-XSPEC}) and the SED fitting from \citetalias{Guolo2025}. The grey shaded region indicates the `effective outer radius' from the time-dependent fit, defined as the radius where the surface density drops to 1\% of its maximum value. The two purple error bars represent the $R_{\rm out}$ estimates from the `high-spin mode' of the 2014/2018 SED fitting presented in \citetalias{Guolo2025}.}
    \label{fig:R_out}
\end{figure}
The derived black hole mass and inclination, $\log(M_{\bullet}/M_{\odot}) = 6.55 \pm 0.15$ and $i = 53^\circ \pm 10^\circ$, are consistent with the range of values inferred in \citetalias{Guolo2025}. In this context, it is important to recall—as briefly noted in \citetalias{Guolo2025}—that the often-assumed black hole mass for \target\ of $M_{\bullet} \sim \mathrm{few} \times 10^5 \ M_{\odot}$, based solely on the central value of the highly uncertain $M_{\bullet}$–$\sigma_*$ relation, is strongly inconsistent with the observed SED of \target, even after accounting for all plausible systematic uncertainties.

As we stressed earlier, a key distinction between the time-independent approach of \citetalias{Guolo2025} and the fully time-dependent model presented here lies in how data from different epochs are treated. In the former, each epoch is analyzed separately, with parameters such as the peak disk temperature ($T_{\rm p}$) and outer disk radius ($R_{\rm out}$) fitted independently for each observation. In contrast, the time-dependent approach employs only four parameters ($r_0$, $t_0$, $t_{\rm visc}$ and $M_{\rm d}$) to self-consistently describe the disk's evolution across all epochs, with all subsequent evolution given by mass and angular momentum conservation. 
A key parameter in such modeling is the ``viscous'' timescale ($t_{\rm visc}$) of the disk, which sets the natural timescale for the evolution of all of the disk’s properties. The late-time behavior of the solutions to Eq.~\ref{eq:fund} all decay at large times, with a scale set by the magnitude of $t_{\rm visc}$. Under a null-stress inner boundary condition \citep[][which we assume and show adequately describes the spectral evolution]{Cannizzo1990,Mummery_Balbus2021}, the key scaling relations are:

 \begin{equation}\label{eq:Lbol_t}
      L_{\rm Bol} \propto (t/t_{\rm visc})^{-n}, 
 \end{equation}
\begin{equation}\label{eq:Md_t}
    M_{\rm d}(t) \propto (t/t_{\rm visc})^{1-n}, 
    \end{equation}
\begin{equation}\label{eq:Tp_t}
    T_{\rm p}(t) \propto (t/t_{\rm visc})^{-n/4},
\end{equation}
\begin{equation}\label{eq:Sig_t}
    \Sigma(r, t) \propto (t/t_{\rm visc})^{-n}, 
\end{equation}
\begin{equation}\label{eq:Rout_t}
    R_{\rm out}(t) \propto (t/t_{\rm visc})^{2n-2}.
\end{equation}
In these expressions $n$ is the one unknown, which is set by one's choice of disk stress parameterization. The dependence on this index is weak, and classical $\alpha$-theory in a gas pressure dominated disk has $n \approx 1.2$, while a $\mu = 0$ disk has $n \approx 1.3$ \citep{Cannizzo1990,Mummery2020,Mummery2021b}. 

The inferred high value of $t_{\rm visc} \approx 2300 \pm 150$ days is physically the primary driver of the unusually gradual evolution of \target's disk, a point we will explore in more detail in \S\ref{sec:long-lived}, particularly in comparison to other TDEs. The preferred disk model solutions indicate that the X-ray emission peaks at epochs close to the discovery time (i.e., $t_{X0} \approx \mathrm{MJD}\,55491$). Naturally, since the peak X-ray luminosity occurs when $\dot{M}(R_{\rm I})$ reaches its maximum, the initial disk formation time is estimated to be $t_0 \sim \mathcal{O}(t_{X0} - t_{\rm visc})$. This places the formation around $\mathrm{MJD}\,54000$, with a scatter of a few hundred days, or approximately 2005–2007.


From Figure~\ref{fig:corner}, we note that the initial disk radius ($r_0$) is the only parameter whose marginalized posterior reaches the boundary of its prior range, i.e., $250 \leq r_0/r_g \leq 300$. This is unsurprising, as constraints on the disk's outer extent come primarily from UV/optical observations—of which only two epochs (2014 and 2018) are available, both occurring at very late times relative to $t_0$. By those epochs, the disk had already undergone substantial expansion, as previously discussed.
Given the absence of early-time UV observations at \textit{HST} resolution, and the fact the disk's rise phase was not captured—even in X-rays—the model's preference for large $r_0$ values at the edge of the prior is a natural consequence of likelihood maximization. In effect, the model selects the largest permissible $r_0$ that allows the disk to be sufficiently extended once UV data becomes available. As noted earlier, $r_0$ should not be interpreted literally, given the simplified nature of the initial conditions in the Green’s function solution to Eq.~\ref{eq:fund}. Instead, what is key is that the model accurately reproduces the disk's physical expansion from a compact initial state.

In a time-independent SED-fitting approach, the disk is assigned a outer radius $R_{\rm out}$, where the flux integral (e.g., Eq.~\ref{eq:flux}; see also \citealt{Guolo_Mummery2025}) is computed from an inner radius out to a finite, well-defined $R_{\rm out}$.However, in a more realistic time-dependent model, the disk surface density does not exhibit a sharp truncation but instead transitions smoothly from a shallow power-law decay to an exponential cutoff—clearly seen in the bottom-right panel of Figure \ref{fig:profiles}. As a result, a precise outer radius is not well defined. Nonetheless, for illustrative and comparative purposes, we estimate an effective outer disk radius as the location where the surface density falls to a small fraction (e.g., 1\%) of its peak value, providing an approximate, time-dependent measure of disk size. In Figure~\ref{fig:R_out}, we compare this estimate with $R_{\rm out}^{2014}$ and $R_{\rm out}^{2018}$, as inferred from the time-independent SED-fitting approach of \citetalias{Guolo2025}; the results are fully consistent, not only in absolute values, but also in the relative expansion from 2014 to 2018.

Lastly, the initial disk mass, $M_{\rm d} =0.4^{+0.15}_{-0.10}$ \msun, is primarily constrained by energy considerations—it represents the mass required in the disk to sustain the model’s luminosity given the radiative efficiency. 
This disk mass implies a minimum disrupted star mass of $M_{\star} > 1\ M_{\odot}$. While this is on the higher end considering an IMF-like star distribution and that $f_d$ (the fraction of the stellar mass, that settle into the disk) is likely $< 1/2$—it remains entirely plausible for a TDE origin. It is also important to note that this long-lived nature of \target, does not mean that the total accreted mass need to be unreasonably high (indeed, we have measured it to be only $\sim0.4 M_\odot$), and it certainly does not imply that \target should be interpreted as an AGN. 

\subsection{Constraints, Assumptions, and Sources of Uncertainty}\label{sec:limitations}

Having used a theoretical framework (\S\ref{sec:theory}) to fit the data (\S\ref{sec:data}) and infer system parameters (\S\ref{sec:results}), we now assess how these results depend on model assumptions, particularly systematic uncertainties and \textit{ad hoc} choices. We aim to identify which parameters are data-driven and which are model-sensitive, and whether their uncertainties are primarily statistical or systematic.

Although our modeling is fully relativistic—both in terms of the disk fluid dynamics and photon propagation, allowing for spin inference and for classical degeneracies with other parameter to be broken, the value of the black hole spin parameter, $a_{\bullet}$, should be interpreted with caution and not taken too literally.
This is because the inferred spin is known to depend sensitively on some of the assumptions. i) The adoption of a \textit{zero-torque} (or null-stress) inner boundary condition; this assumption, following from \citet{Shakura1973}, is made for modeling simplicity and is commonly employed in most analytical treatment of accretion disk, and in all, but one \citep[\texttt{fullkerr,}][]{Mummery24Plunge} of those available in fitting packages. However, it is known to be physically inaccurate, as shown by GRMHD simulations \citep[e.g.,][]{Noble2011,Rule2025}, and by detailed modeling of X-ray binary spectra \citep{Mummery24Plunge,Mummery2025}. ii) The choice of color-correction prescription. While the use of a color-correction factor, rather than assuming none, is strongly supported by radiative transfer simulations of disk atmospheres \citep[e.g.,][]{davis2006}, the commonly used analytic prescriptions \citep[e.g.,][]{Hubeny2001,Done2012} still represent a simplification of the underlying physics. In practice, uncertainties in the color-correction factor $f_c$ often dominate the error budget in black hole spin measurements \citep{SalvesenMiller2021}.

As discussed throughout this paper, the initial condition employed in our model—namely, an initial ring of material—is an artificial simplification, albeit a necessary one for analytical treatment. In reality, the disk formation process is highly complex \citep{Lu2020,Bonnerot2020,Steinberg2024}, and observational evidence suggests this simplifications can fail in describing the X-ray spectra of TDEs during the rise of their X-ray light curves \citep{Guolo2024}. 
Consequently, neither $r_0$ nor $t_0$ should be taken literally. Furthermore, the detailed shape of the rising X-ray light curve should not be interpreted as a precise model prediction. 
Nevertheless, it is important to emphasize that, in the case of \target, at all times where data are available, the source can be well described across all wavelengths by an evolving thin accretion disk.\footnote{Direct disk emission alone is insufficient to reproduce the early-time bright optical/UV emission commonly observed in optically selected TDEs. This component, if present, was never observed in \target\ and is therefore not modeled here.}

At this point, it is well established that $N_H$, $E(B-V)$, $M_{\bullet}$, and $i$—as well as other parameters such as peak disk temperature ($T_p$) and $R_{\rm out}$, which are `observables' rather than free parameters in \S\ref{sec:theory}—can be inferred from the multi-wavelength continuum emission of accreting black holes. While degeneracies exist between these parameters, they can often be broken if high-quality, multi-epoch data is available. 
For instance, $M_{\bullet}$ can only be precisely inferred from continuum fitting if some information, even if upper-limit, on the thermal emission from the inner portion of the disk is directly observed in the X-rays (which, e.g., is not the case for most AGN, given their higher black hole mass and the Comptonization effects present). On the other hand, the inference of $i$, as well as $R_{\rm out}/r_g$, requires a relativistic framework and simultaneous X-ray and UV/optical data \citep[see relevant discussion in][]{Guolo_Mummery2025}.

Models for QPE flares that invoke collisions between an orbiting object and the TDE disk are sensitive to the surface density of the disk at the collision radius (this sets how much material is collided with, and the gross energy scales of the collision itself). Unlike in steady-state models, in a time-dependent disk model framework the amplitude of the surface density and disk mass can be constrained from multi-epoch data. The reason for this is that the temperature (in principle an observable) is defined by the product $W^r_\phi \Sigma$ (see equation \ref{eq:t}), but the evolutionary timescale of the disk (also in principle an observable, see \S\ref{sec:long-lived}) is sensitive to only $W^r_\phi$, meaning this degeneracy can be broken. It is this second timescale measurement which is inaccessible to steady-state approaches. There is some subtlety to the value of $\Sigma(r)$ related to the radius dependence of $W^r_\phi$ \citep{Lynden-Bell1974},  but this at most introduces order $1$ factors, and cannot change the measured scale found here\footnote{To really stress this point, an equivalent way of thinking about this is that the amplitude of the surface density is really only dependent on the total mass of the disk, which is observationally related to the total energy observed over long timescales.}.

While the absolute value of the surface density radial profile, $\Sigma(r)$, at radii $r_I \ll r \ll R_{\rm out}$ is somewhat dependent on the stress parametrization (i.e., the value of $\mu$ in Eq.~\ref{eq:stressdef}), an important property---to be recalled in \S\ref{sec:qpes}---is that the \emph{ratio} of surface densities at a fixed radius $r_i$ at two different times $t_1$ and $t_2$, namely $\Sigma(r_i, t_1)/\Sigma(r_i, t_2)$, is \textit{independent} of the choice of $\mu$. To see this, we consider the ratio of two realizations of Eq.~\ref{eq:t}:

\begin{equation}\label{eq:sigma_T}
    \frac{T^4(r_i, t_1)}{T^4(r_i, t_2)} = \frac{W^r_\phi(r_i, t_1)}{W^r_\phi(r_i, t_2)}\frac{\Sigma(r_i, t_1)}{\Sigma(r_i, t_2)} = \frac{\Sigma(r_i, t_1)}{\Sigma(r_i, t_2)} , \quad \forall \mu,
\end{equation}

\noindent where $T(r_i, t)$ is the local disk surface temperature at $r_i$\footnote{In principle, and in classical $\alpha$ models, $W^r_\phi$ can depend on the disk density to some (typically small) positive power. This will reduce the change in $\Sigma$ for a given change in $T$, something which we shall show is the opposite dependence for what collisional QPE models require. For QPE model discussions therefore this is a conservative assumption.}. Moreover, since $T(r_i) \propto T_p \left( {r_i}/{r_p} \right)^{-3/4}$, at $r \geq r_p$, for all choices of $W^r_\phi$ 
 - where $r_p$ is defined as the radius at which the temperature peaks, i.e., $T_p = T(r_p)$ - we can equivalently express Eq.~\ref{eq:sigma_T} in terms of $T_p$ as:

\begin{equation}\label{eq:sigma_T2}
    \frac{T_p^4(t_1)}{T_p^4(t_2)} = \frac{\Sigma(r_i, t_1)}{\Sigma(r_i, t_2)}, \quad \forall \mu.
\end{equation}

\noindent Since it is well established that $T_p$ can be directly inferred from the X-ray spectrum, and its evolution $T_p(t_1)/T_p(t_2)$ can be constrained from multi-epoch observations, it follows that the ratio $\Sigma(r_i, t_1)/\Sigma(r_i, t_2)$ can also be observationally inferred—independently of the stress parametrization assumed and at the same precision as $T_p^4(t_1)/T_p^4(t_2)$.

\subsection{Implications for QPEs Models}\label{sec:qpes}
There is, by now definitive evidence that QPEs are linked to the presence of an accretion disk. Most sources show a persistent, but evolving, X-ray emission between eruptions, and UV/optical counterparts of this have recently been detected \citep{Nicholl2024,Wevers2025,Guolo2025,Chakraborty25} in some of the sources. Their X-ray emission is soft and thermal, indistinguishable from those of typical X-ray detected TDEs, while they lack the hard power-law coronal component ubiquitous to AGN;  none of the sources display a persistent and virialized broad-line region. Where sufficient data exist, the inferred disk sizes are compact, again consistent with a TDE origin \citep[e.g.,][]{Nicholl2024,Wevers2025,Guolo2025,Chakraborty25}. QPEs have now been discovered in two typical optically selected TDEs—AT2019qiz and AT2022upj \citep{Nicholl2024, Chakraborty25}—and likely in at least two more: AT2019vcb \citep{Quintin2023,Bykov2024} if its X-ray flares are interpreted as QPEs, and \target, if it is considered a TDE, as we argue in \citetalias{Guolo2025} and in Section~\ref{sec:long-lived}.

Given that QPEs are related to the presence of an accretion disk - specific a TDE-fed disk - the properties of the eruptions most likely correlate with those of the disk. Two main model classes have been proposed and could explain these connections: one invoking disk instabilities \citep[e.g.,][]{Pan2022, Pan2023, Kaur2023}, and another involving interactions between an orbiting body and the disk. In particular, models based on orbiter-disk collisions \citep[e.g.,][]{Linial2023, Franchini2023, Yao2024} have gained popularity for naturally explaining the QPE–TDE connection and given the fact that the appearance of eruptions correctly match the disk surface density evolution observed in AT2019qiz \citep{Nicholl2024}. 

In \citetalias{Guolo2025}, we employed a relativistic, though time-independent, accretion disk model to simultaneously fit the X-ray and UV spectra of \target\ at two epochs: 2014 and 2018. While QPEs were discovered in the 2018 observations, they were intrinsically\footnote{The absence of X-ray eruption in the 2014 data, is intrinsic and can not be explained the the fact the disk was hotter in 2014 than in 2018, as shown in both Appendix A of \citetalias{Guolo2025} and Appendix C of \citet{Miniutti2023_noqpe}. If the eruptions are present in 2014, they need to be have substantially lower temperature then they had in 2018, see Appendix \ref{app:T_qpe}.} absent in 2014 (see inset in Fig.~\ref{fig:lc+SED}). This modeling already enabled more precise constraints—compared to previous X-ray-only analyses—on several system parameters relevant to QPE models, particularly the Eddington-normalized accretion rate and the outer disk radius.

We summarize the key findings from that study, which are reproduced here for a more detailed disk model: (i) No existing (i.e., published) disk instability model can account for the stability of the disk in 2014 (no QPEs) and its instability in 2018 (QPEs present). In fact, the Eddington ratio was higher in 2014 than in 2018, contrary to the expectations of models that predict a clear (in)stability criterion based on accretion rate \citep{Kaur2023}, with the opposite dependence (i.e., less stable at higher Eddington ratios). (ii) Although the 2018 disk was sufficiently extended to allow for orbiter/disk interactions (i.e., the inferred orbital radius $r_{\rm orb}$ was smaller than $R_{\rm out}$), the 2014 disk was already comparably large—yet QPEs were absent.

\begin{figure}
	\centering
	\includegraphics[width=\columnwidth]{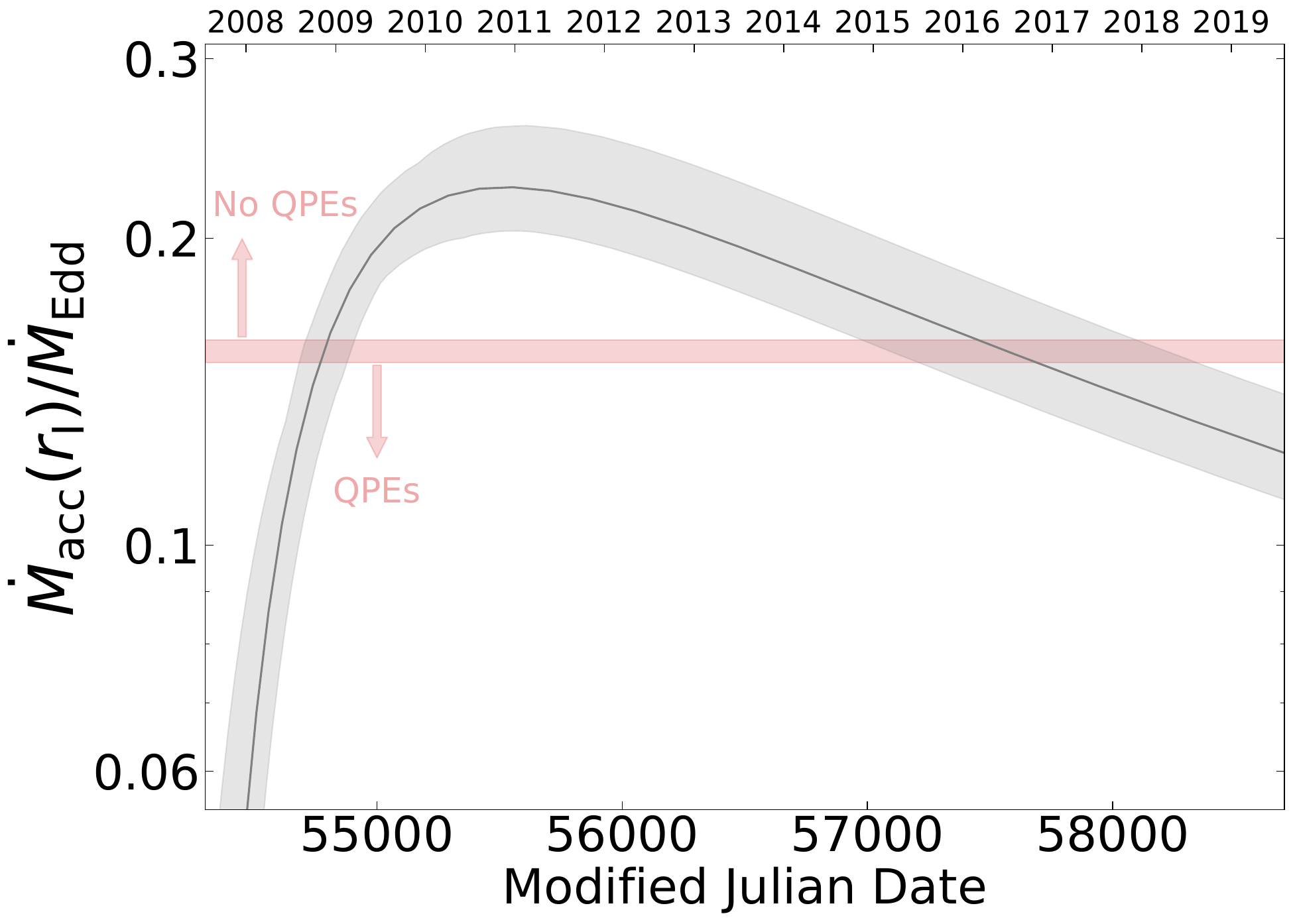}
 
	\caption{Inferred Eddington-normalized mass accretion rate of \target as a function of time. The red line illustrate the apparent transition value between presence (absence) of QPE at low (high) accretion rates. Figure inspired from \citet{Miniutti2023_noqpe}. The observed behavior is the opposite of predicted by disk instability models.}

    \label{fig:kaur}
\end{figure}

Before discussing implications of our time-dependent disk fitting for orbiter-collision models, we briefly revisit the challenges disk instability models face in explaining QPEs in \target. As detailed in \citetalias{Guolo2025}, models that define an explicit instability criterion \citep[e.g.,][]{Kaur2023}\footnote{Many proposed instability models lack a well-defined triggering criterion and therefore cannot account for the appearance and disappearance of QPEs in \target, or the lack of QPE in most TDE disks.} typically predict that eruptions should occur when the disk’s Eddington ratio exceeds a critical threshold, $\lambda_{\rm Edd} \geq \lambda_{\rm Edd,crit}$. Below this threshold, the disk should be stable.
In \target, however, we observe the exact opposite: as shown in Fig.~\ref{fig:kaur}, QPEs are present at low $\lambda_{\rm Edd}$ and absent at high $\lambda_{\rm Edd}$. This inversion is consistent with the findings of \citet{Miniutti2023_noqpe}, who first noted this accretion-rate dependence, and extended it to post-2020 data. While our inferred transition $\lambda_{\rm Edd}$ differs slightly from theirs, as is to be expected given differences in how $M_{\bullet}$ and accretion rates are estimated, the trend is identical. 

An alternative possibility in the context of disk instabilities, proposed by \citet{Pan2023}, is that eruptions were already present in 2010 and 2014, but with a sufficiently long recurrence period ($P_{\rm QPE}$) such that no flares were caught during the X-ray observations. In particular, the 2014 dataset would require $P_{\rm QPE}(2014) \gtrsim 25$ hours (the duration of that observation). However, we note that the flare duration does not evolve with accretion rate during the epochs when eruptions are consistently detected, despite significant variations in Eddington ratio. Between 2014 and 2018, the Eddington ratio increased only modestly (by a factor of 1.36), so explaining a change in $P_{\rm QPE}$ by a factor $\gtrsim 3$ (from $\gtrsim 25$ hr to $\sim$9 hr) would require an unusually steep dependence, $P_{\rm QPE} \propto \lambda_{\rm Edd}^{\sim 4.2}$. Such a scaling would also predict a measurable change between 2018 and 2019, when $\lambda_{\rm Edd}$ declined by $\sim$5\%, implying a reduction in period from $\sim$9 hr to $\sim$7.6 hr. No such change is observed \citep[Fig. 3 in][]{Miniutti2023_longterm}. Indeed, none of the known QPE sources show evidence for a correlation between recurrence period and accretion rate.

Finally, we note that all existing disk instability models for QPEs \citep[e.g.,][]{Pan2022,Pan2023,Sniegowska2023,Kaur2023} begin with steady-state disk models and introduce $\alpha$-model instabilities on top. Since TDE disks are by definition time-dependent, not in a steady-state, and it is likely that the $\alpha$-model is a gross simplification of the physics of real MHD turbulence, it is not clear whether instabilities (of some form) can be ruled out from the data, or if real instabilities are simply more complex than current simple models.   

We now discuss collisional models for QPEs.  A key feature of the \citet{Linial2023}'s disk-orbiter framework (building on the supernova shock breakout models of \citealt{Nakar10}) offers a potentially natural  explanation to the lack of QPEs in 2014, even with an sufficiently extended disk. In this model, disk-orbiter collisions produce detectable X-ray radiation only if the shocked gas is in the photon-starved regime\footnote{Photon starvation is the term used for when there are an insufficient number density of photons to carry the energy in the collision away with a black body spectrum, meaning the escaping photons must have a harder spectrum for energy to be conserved.}. If the disk's surface density at the interaction point is sufficiently high, the post-shock gas thermalizes with the radiation field efficiently and radiates at the post-shock gas temperature $T_{\rm sh}$, which for typical conditions is $\sim 10$~eV—too soft to be observed in X-rays \citep{Linial2023, Vurm2024}. Thus, the absence of X-ray QPEs in 2014, despite a large disk, could be attributed to a higher surface density at the radius of the invoked orbiter ($r_{\rm orb}$)\footnote{In the case of circular orbits, $r_{\rm orb}$ is related to the recurrence time of the QPEs ($P_{\rm QPE}$) by the relation $P_{\rm QPE} = \pi \sqrt{r_{\rm orb}^3 / G M_{\bullet}}$, for \target, which $P_{\rm QPE} =9\pm1$h and log($M_{\bullet}$/\msun) $= 6.55 \pm 0.15$, this gives $50 \leq r_{\rm orb}/r_g \leq 100$.}, suppressing X-ray emission. The subsequent emergence of QPEs in 2018 would then require a significant drop in $\Sigma(r_{\rm orb})$ over time. With the time-dependent disk modeling applied in this work - which is physically the same as applied in \citet{Nicholl2024,Chakraborty25} - we are now able to test this possibility directly.

\begin{figure}
	\centering
	\includegraphics[width=\columnwidth]{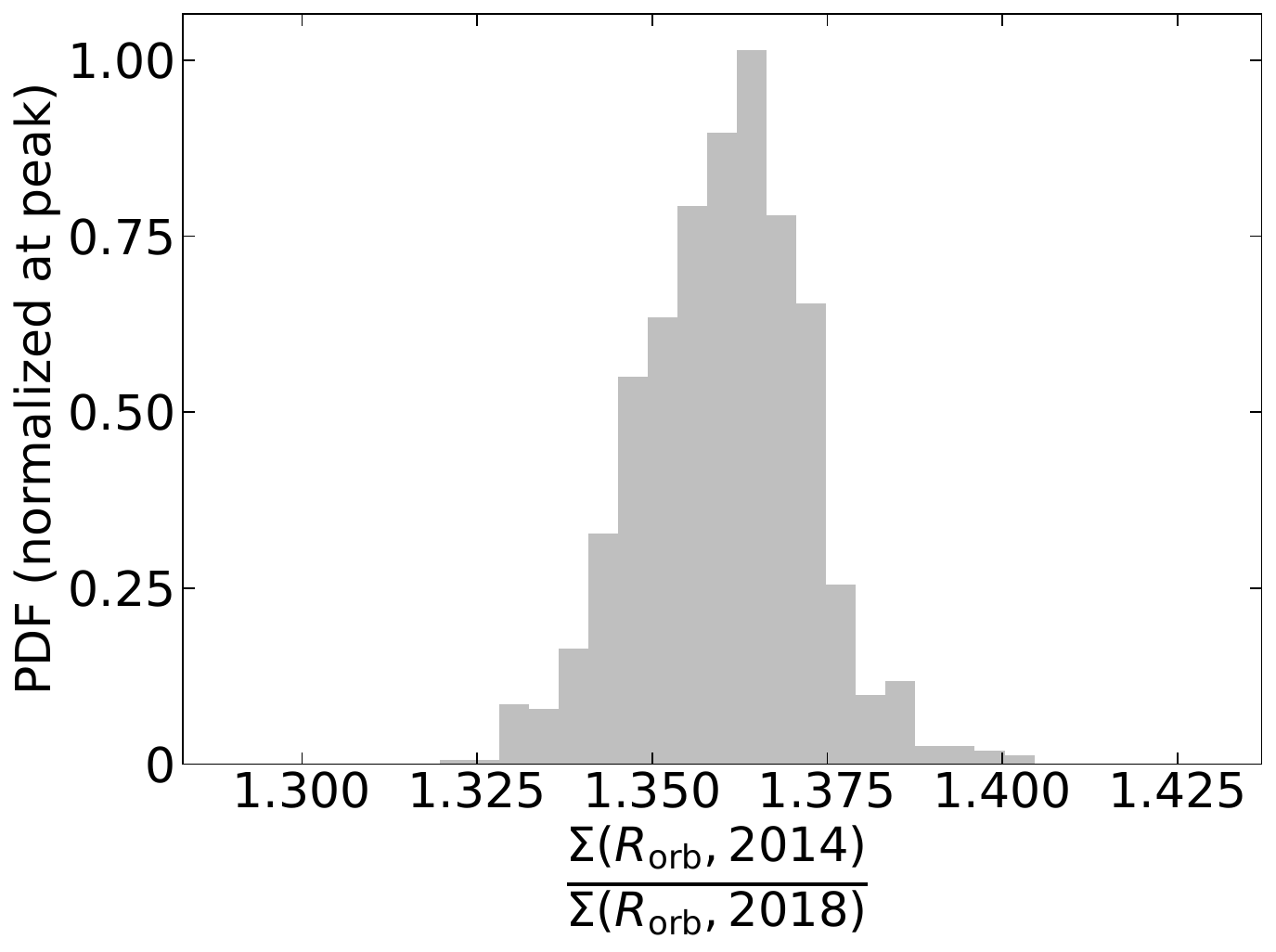}
 
	\caption{Probability density function of the change in disk's mass surface density at the position of the proposed orbiter (for disk-collision QPE models), from 2014 to 2018.}
    \label{fig:sigma_hist}
\end{figure}

In Fig.~\ref{fig:sigma_hist}, we show that the probability distribution function (PDF) of the change in surface density in the orbiter radius: $\Sigma(r_{\rm orb}, 2014)$ was $\sim 36 \%$ higher than $\Sigma(r_{\rm orb}, 2018)$. Here, we recall our results from \S\ref{sec:limitations} that ${\Sigma(r_i, 2014) }/{\Sigma(r_i, 2018) }$ is physically related to ${T_p(2014) }/{T_p(2018)}$, which is directly probed from the X-ray spectra, and it is not dependent on our parametrization of the stress tensor.

In the \citet{Linial2023}'s framework the shock-heated material after the collision emits at an observable QPE temperature ($T_{\rm obs}^{\rm qpe}$), which depends on whether or not the shocked gas reaches thermal equilibrium or if it is in the photon-starved regime, and is given by: 

\begin{equation}\label{eq:regime}
    T_{\rm obs}^{\rm qpe} = T_{\rm sh} \times  \left\{\begin{matrix}
1, & \eta < 1 & {\rm (thermal \  equilibrium)}\\
\eta^2, & \eta > 1  & {\rm (photon{\rm -}starved)}\\
\end{matrix}\right. 
\end{equation}

\noindent where $\eta$ is the dimensionless photon production efficiency coefficient \citep{Nakar10}. While the analytical expression for both  $T_{sh}$ and $\eta$ are complicated (but see them in \citet{Linial2023} and \citet{Mummery2025}), important to our purpose, is that it can be shown \citep{Mummery2025}, that they depend on $\Sigma(r_{\rm orb})$ as  $T_{\rm sh} \propto \Sigma^{-1/4}$ and $\eta \propto \Sigma^{-9/8}$, such that $T_{\rm obs}^{\rm qpe} \propto \Sigma^{-5/2}$. The other parameters which they depend on are not expected to vary between epochs. For our interests, the following equation is then useful:

\begin{equation}\label{eq:test_LN23}
\frac{T_{\rm obs}^{\rm qpe} (t_1)}{T_{\rm obs}^{\rm qpe} (t_2)} = \left ( \frac{\Sigma(r_{\rm orb}, t_2)}{\Sigma(r_{\rm orb},t_1)} \right )^{5/2} = \left(\frac{T_{p} (t_2)}{T_{p} (t_1)} \right)^{10} .
\end{equation}

\begin{figure}
	\centering
	\includegraphics[width=\columnwidth]{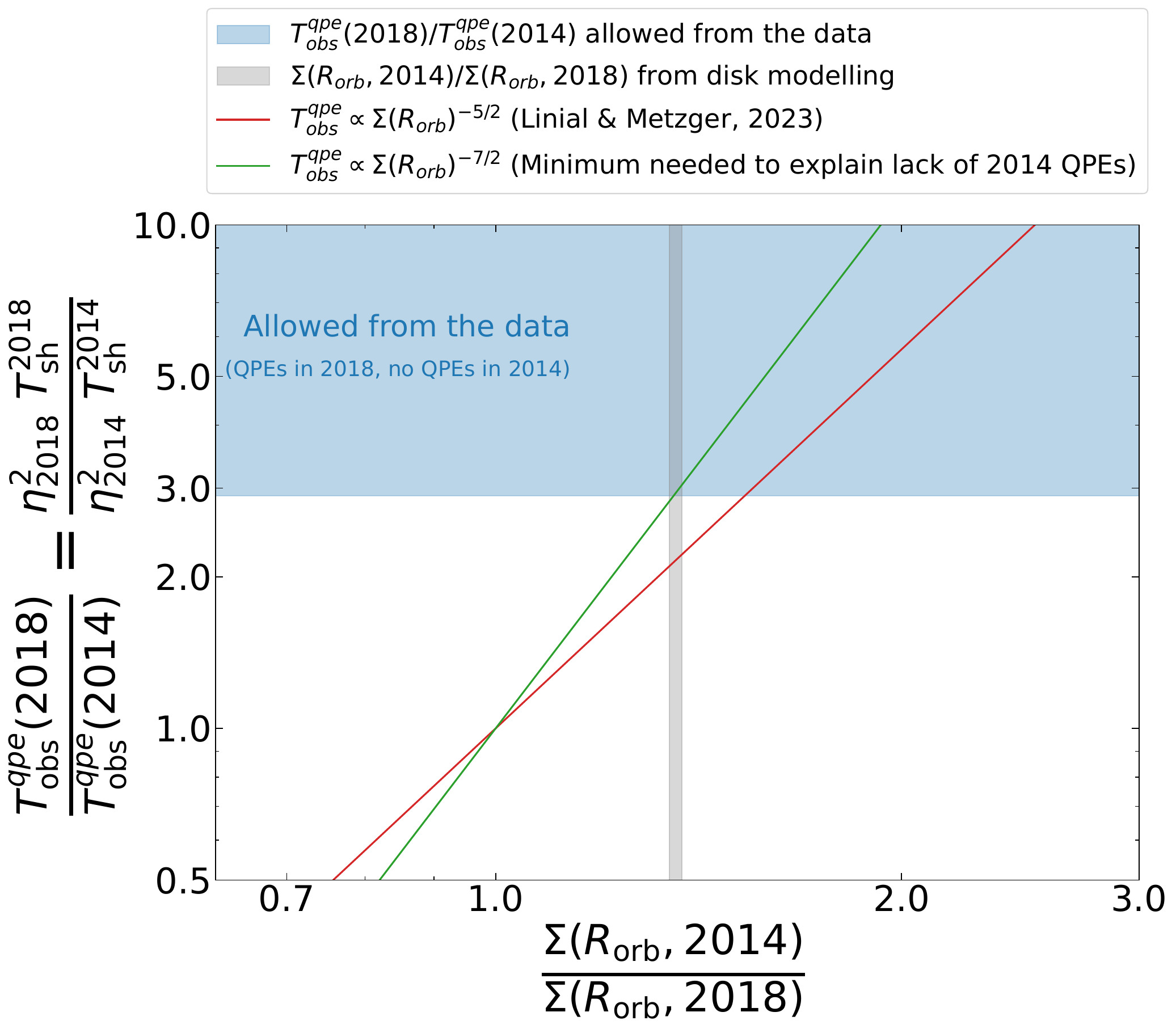}
 
	\caption{Summary of the results for the our test of disk-orbiter collision model for QPEs. The y-axis shows the change in QPE temperature between 2014 and 2018, with the blue region representing the range allowed by the data (i.e., QPEs detected in 2018 but not in 2014). The x-axis shows the change in disk surface density at the orbiter radius ($\Sigma(r_{\rm orb})$), with the gray shaded region indicating the range inferred from our time-dependent disk modeling. The red line shows the prediction from the model of \citet{Linial2023} (Equations~\ref{eq:regime} and~\ref{eq:test_LN23}). The figure illustrates that a change in  $T_{\rm obs}^{\rm qpe}$ towards undetectable values does not seem to be able to account for the absence of QPEs in 2014 and their appearance in 2018, as the change in $\Sigma(r_{\rm orb})$ was insufficient. See text for a detailed discussion.
}
    \label{fig:LM23}
\end{figure}

Given that QPEs were detected at their peak in the 2018 observations with an observed temperature of k\( T_{\rm obs}^{\rm qpe}(2018) \sim 105\,\mathrm{eV} \), and that no X-ray QPEs were observed in 2014, the data (see Appendix~\ref{app:T_qpe}) imply that the QPEs in 2014 must have had \( T_{\rm obs}^{\rm qpe}(2014) \lesssim 36\,\mathrm{eV} \) (99\% credible interval), in order for the eruption not to be detected above the disk emission at the eruption's peaks, therefore the observed temperature ratio \( T_{\rm obs}^{\rm qpe}(2018) / T_{\rm obs}^{\rm qpe}(2014) \gtrsim 2.9 \) is an observational constraint.
With both this temperature ratio and the surface density ratio \( \Sigma(r_{\rm orb}, 2014) / \Sigma(r_{\rm orb}, 2018) \approx 1.36 \) measured from the data, Fig.~\ref{fig:LM23} illustrates the difficulty faced by the model of \citet{Linial2023}. From Eq.~\ref{eq:test_LN23}, this surface density variation would only predict a maximum temperature change of \( \left( \Sigma(r_{\rm orb}, 2014) / \Sigma(r_{\rm orb}, 2018) \right)^{5/2} \approx 2.2 \), below the observed value $\gtrsim 2.9$. 

This indicates that a transition from thermalized to photon-starved shocked gas, or a simple change in  $T_{\rm obs}^{\rm qpe}$ towards  unobservable values, i.e. given by  $T_{\rm obs}^{\rm qpe} \propto \Sigma(r_{\rm orb})^{-5/2}$, as proposed by \citet{Linial2023} following \citet{Nakar10}'s prescription (Eq.~\ref{eq:regime}), cannot simply explain the absence of X-ray QPEs in 2014 and their emergence in 2018. For instance, in this framework, reproducing the observed evolution would require a relation of the form $T_{\rm obs}^{\rm qpe} \propto \Sigma(r_{\rm orb})^{-\gamma}$ with $\gamma \gtrsim 7/2$. This represents a significant deviation from the expected $\gamma = 5/2$, and at present, it is unclear what additional physical processes could produce this steeper dependence.
However, such step dependence, which would result in $T_{\rm obs}^{\rm qpe} (t_1) / T_{\rm obs}^{\rm qpe} (t_2) $ = $(T_{p} (t_2) / T_{p} (t_1))^{14}$! would likely be already inconsistent with the only mild evolution of $T_{\rm obs}^{\rm qpe}$ in GSN observed in \citet[][see their Fig~B.2]{Miniutti2023_noqpe}. It is important to emphasize that this diagnostic is independent of the nature of the orbiter—it applies whether the colliding object is a star, a compact object, or an EMRI puffed up to fill its Hill sphere \citep[e.g.,][]{Yao2024}, as shown in \citet{Mummery2025}.

To our knowledge, combining \citetalias{Guolo2025} and this section, we have now explored all currently available explanations in the literature that could account for the absence of QPEs in \target\ prior to 2018, and their emergence at such late times, given the observed and constrained physical properties and evolution of the `quiescent' disk. 

In the context of disk-eruption co-evolution, it is also relevant to comment on the results of \citet{Chakraborty2024} for eRO-QPE1 \citep{Arcodia2021}, in which the authors report a decreasing $T^{\rm qpe}_{\rm obs}$ with decreasing X-ray luminosity ($L_{\rm disk}$) and cooling of the `quiescent' disk emission. We note, that this trend is in the \textit{opposite} direction to what is expected from the collision scenario, if the disk has a finite mass supply (i.e., is a TDE-fed disk). The reason for that, is that in such disks, $L_{\rm disk} \propto T_{\rm p}^4  \propto \dot {M}(r_{\rm I}) \propto \Sigma(r_{\rm p}) \propto \Sigma(r_{\rm orb})$ \citep[see detailed discussion in][]{Mummery2025}, while the collision model predicts $T_{\rm obs}^{\rm qpe} \propto \Sigma(r_{\rm orb})^{-5/2}$. Therefore, within this framework, $T^{\rm qpe}_{\rm obs}$ should anti-correlate with $L_{\rm disk}$, not correlate positively as observed for eRO-QPE1.

While these findings cast some doubt on the viability of orbiter--disk collision models as an explanation for QPEs---at least in their simplest form---we cannot rule out the possibility that the relevant physics is more complex. Just as disk instabilities may involve processes beyond the standard $\alpha$-viscosity prescription, the radiative transfer from post-collision shocked debris may not be well described by simple analytical models. Numerical simulations \citep[e.g.,][]{Vurm2024} will be essential in addressing this question, and may benefit from the empirically constrained disk properties of GSN 069 presented in this work.

More broadly, beyond the specific case of \target, we emphasize that while the complex timing behavior of QPEs—often significantly deviating from quasi-periodicity \citep[e.g.,][]{Arcodia2022,Chakraborty2024,Chakraborty25}—can be reconciled by introducing additional degrees of freedom, such as various forms of precession \citep[e.g.,][]{Franchini2023,Chakraborty2024,Chakraborty25} or even a third body in the system \citep{Miniutti2025}, other key observational features are harder to reproduce in a self-consistent manner. For example, the appearance and disappearance of eruptions \citep[e.g.,][]{Miniutti2023_noqpe}, their apparently universal late-time emergence \citep[e.g.,][]{Nicholl2024,Chakraborty25}, and more specifically the absence of eruptions in \target\ in 2014, seems to be harder to explain by existing orbiter-based models, and also by any currently available QPE model.

We believe that the absence of QPEs in \target\  in 2014, and the their appearance in 2018, is perhaps the best posed of these open problems, as we have good constraints on the disk properties at both times. We believe that this constitutes a major open challenge in the field and should be a central focus of future theoretical studies.

\subsection{Discussion of the 2020-2022 rebrightening event}\label{sec:2020}

\begin{figure}[]
	\centering
	\includegraphics[width=0.9\columnwidth]{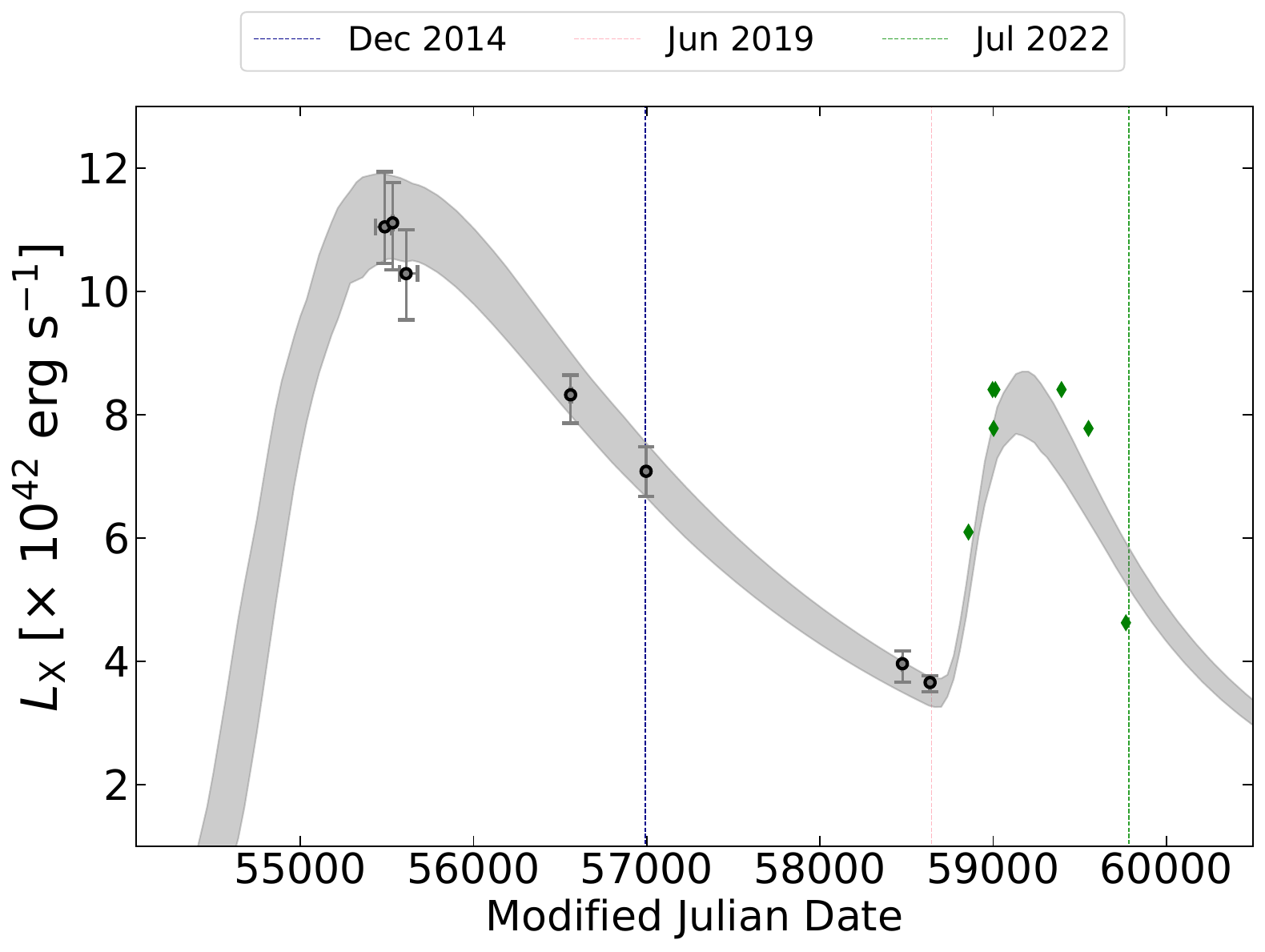}\\
    \includegraphics[width=0.9\columnwidth]{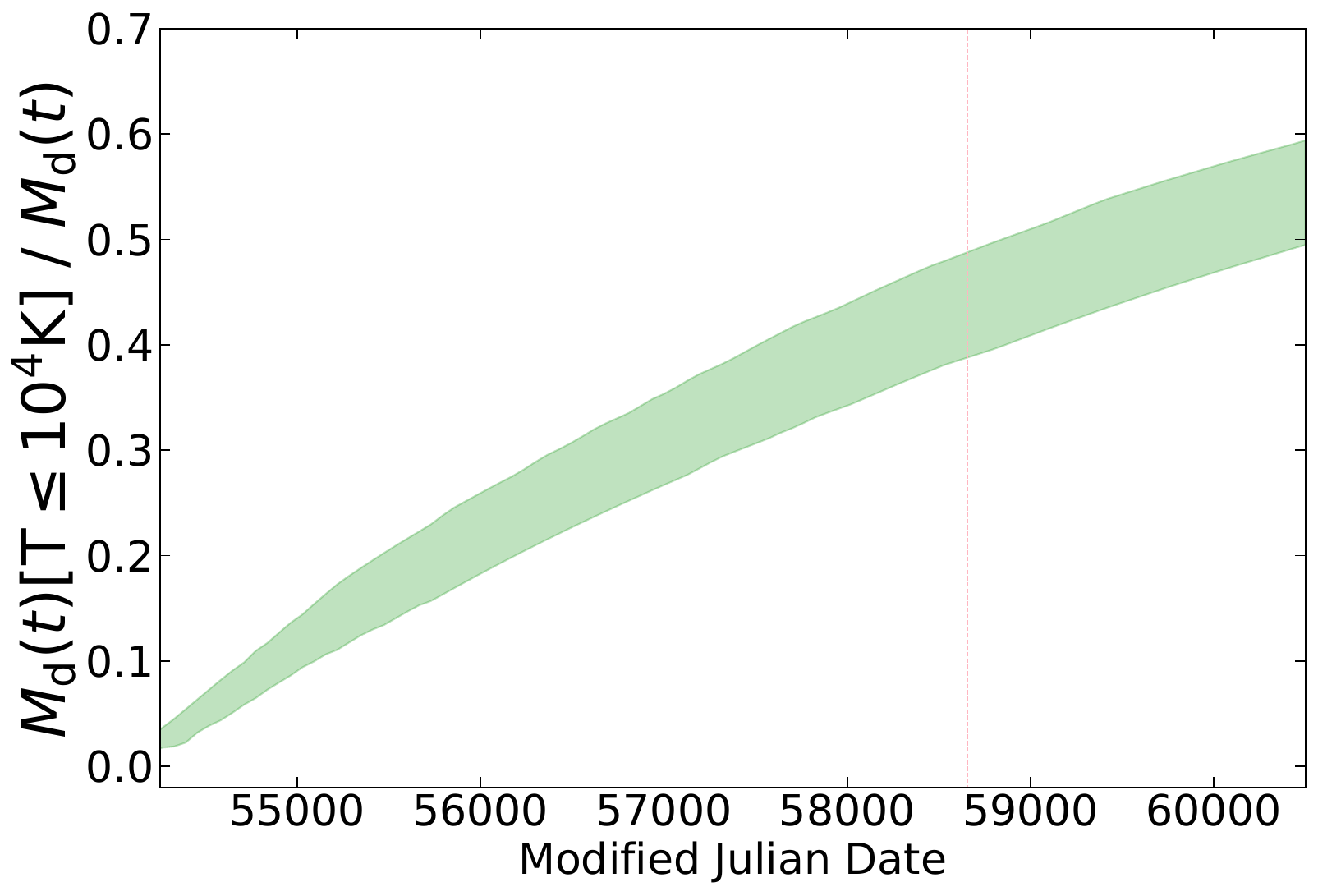}
 
	\caption{Study of the 2020--2022 rebrightening event. \textit{Top:} Illustrative addition of a second \texttt{FitTeD} solution to reproduce the flare. \textit{This is not a fit}—neither to the flare’s luminosity nor its spectra. \textit{The green points were simply scaled linearly from the count rate}, without accounting for spectral evolution or uncertainties. Shown for illustrative purposes only. \textit{Bottom:} Fraction of the remaining disk mass with local temperatures $T \leq 10^4\,\mathrm{K}$, as a function of time. This is a relevant quantity for the Hydrogen Ionization Instability, which can only be calculated with a time-dependent model of the disk. Vertical colored lines mark key timescales in the source’s evolution; see text for discussion.
 }
    \label{fig:2020_flare}
\end{figure}

Throughout this paper, we have restricted our analysis to data spanning from \target's discovery through to late 2019. The rationale behind this choice is the emergence of a second X-ray flare event between 2020 and 2022, as reported by \citet{Miniutti2023_longterm}. While our main conclusions—particularly those discussed in \S\ref{sec:qpes} and \S\ref{sec:long-lived}—are not dependent on the presence or properties of this rebrightening event, it is still worthwhile to briefly comment on this second flare in the context of time-dependent accretion theory.

This rebrightening has been interpreted in two main ways: (i) as the addition of more debris to the disk from the return of a partially disrupted star, or (ii) as a disk instability arising from the co-evolution of the disk-orbiter, typically assumed to be driven by the addition of material to the disk owing to collisions between the disk and an invoked EMRI \citep[in the context of collision-driven QPE models, e.g.,][]{Linial2024}. However, these interpretations—and the corresponding analyses—typically rely on one or both of the following assumptions: (i) that the X-ray luminosity traces the fallback rate, and (ii) that the disk structure is that of a steady-state configuration, with an instability developing on top of it. We can however improve/extend on both of these analyses in this work. 

The observed behavior of this second flare can be summarized as follows: it begins to rise slightly after the time of the 4th XMM spectrum $(t_{X4})$ (i.e., in late 2019), peaks around 2021—reaching a luminosity comparable to that observed in 2014—and then decays back to a level similar to the 2019 luminosity by late 2022.

Interestingly, while large scale disk instabilities and the return of a stellar remanent are completely distinct physical scenarios, the subsequent evolution of both classical disk instability models and the addition of extra disk material can, to leading order, be treated in a similar fashion within {\tt FitTeD}. The reason for this is the following. Within the instability scenario, a thermal instability occurs at some radial scale $r_{\rm inst}$, leading to run-away heating (in the instance of a large scale X-ray flare, although runaway cooling is also in principle possible). Most models of the turbulent stress in accretion flows lead to a rapid elevation of $W^r_\phi$ on this unstable hot branch \citep[this, for example, is common in the disk instability model for outbursts in the Cataclysmic Variable community,][for a review]{Hameury2020}. 

Inspection of equation \ref{eq:fund} shows that the evolution of the disk surface density $\Sigma(r, t)$ is driven by radial gradients in the product $W^r_\phi \Sigma$. Thus, the addition of extra material to the disk from a returning partially-disrupted stellar remanent (increasing $\Sigma$ at a certain radial scale), or a disk thermal instability (increasing $W^r_\phi$ at a certain radial scale), can both be treated approximately within the {\tt FitTeD} framework by adding additional disk solutions on top of the relaxed late-time disk structure. 

The interpretation of the resulting additional solution of course differs in the two scenarios. The disk mass parameter in particular corresponds to the extra material added to the flow in the returning remanent scenario, while it corresponds to the amount of the {\it original} disk mass which is ``dragged'' by the heating wave moving through the disk (i.e., the mass of the original disk which is in the disk region which went unstable) in the instability scenario. 

Given the inherent uncertainties in modeling such distinct scenarios, we perform a less rigorous analysis of the subsequent second-flare disk evolution in this section (when compared to the full SED fitting of the first flare), aiming to provide only a qualitative analysis of the second flare.  

To set the scene, the initial (i.e., for the first flare) disk mass was $M_d (t_0) \approx 0.4 M_\odot$ (see e.g., section \ref{sec:results}), and by the time of the second flare an amount $\Delta M_{\rm acc}(t_{2019}) = \int^{t_{2019}}\dot M(r_I, t')\, {\rm d}t' \approx M_d(t_0)/4$ had been accreted, leaving $M_d (t_{2019}) \approx 3 M_d(t_0)/4  = 0.3 M_\odot $ remaining in the disk. We then introduce a second {\tt FitTeD} solution, allowing only the additional mass $M_{d, 2}$, the second start time $t_{0, 2}$ and the second viscous timescale $t_{\rm visc,2}$ to be free parameters\footnote{i.e., we do not allow the black hole mass or spin, the disk-observer inclination, the radial scale of the added mass, nor $N_H$ to vary.}. We then perform a simple illustrative fit to the light curve of the second X-ray flare, assuming that the temperature evolution equation can be treated linearly (i.e., $L_X(t) = L_X(T_1(t) + T_2(t))$, where $T_1, T_2$ are the temperature profiles of the two {\tt FitTeD} solutions). 

The  data for the second X-ray flare are shown by green data points in Figure \ref{fig:2020_flare}. These luminosities are simply \textit{linearly scaled} count rates, and we do not account for the spectral evolution of the source, nor do we propagate uncertainties from either the model or the data. We stress that this solution is supposed to be {\it illustrative}, and not a robust fit given the various uncertainties.

We find that no value $M_{d,2}$  provides a reasonable solution of the second flare if we enforce $t_{\rm visc,2} = t_{\rm visc}$\footnote{This remains the case if we allow $r_0$ to vary for the second flare.}. This means that the second flare propagates on a fundamentally different timescale from the first. If we allow $t_{\rm visc,2}$ to be free, then we can reasonably well describe the data. 

The solution displayed  in Figure \ref{fig:2020_flare} has the following parameters $t_{0,2} -  t_0 \approx 13$ years,  $M_{d,2}(t_{0,2}) \approx M_d(t_0)/4= 0.1M_\odot$ (this is the  mass in the second {\tt FitTeD} solution, not the accreted mass in the three years between 2019 and 2022), and $t_{\rm visc,2} \approx t_{\rm visc}/2$. We do not provide uncertainties on these values, owing to the simplifications involved. 

We see that the addition of debris from a returning remanent remains a plausible solution to the problem, although it is not clear what would result in the change in the viscous timescale of the disk. We also note that a repeating partial TDE requires a somewhat fine tuned setup, as the original mass content ($M_d \sim 0.4 M_\odot$) is rather high for a stripped mass content which leaves behind a surviving remanent. 

It is not clear if the addition of mass stripped from the EMRI can induce a thermal instability. The surface density near the collision radius dropped from $\Sigma \sim 10^5$ kg/m$^2$ near peak luminosity to $\Sigma \sim 5 \times 10^4$ kg/m$^2$ in 2019. To move the local disk surface density outside of the range already spanned by the disk solution (which is presumably required to induce a thermal instability not seen in the initial evolution) would require adding at least $\Delta M_{\rm emri} \sim \pi r_{\rm orb}^2 \Sigma_{\rm add} \sim 0.2M_\odot (\Sigma_{\rm add}/10^6$kg\,m$^{-2}$), where $r_{\rm orb} = (GM_\bullet)^{1/3} (T_{\rm QPE}/\pi)^{2/3} \approx 3.4\times10^{11}$ meters for our best fitting black hole mass. Assuming a collision every 9 hours for $\sim 6$ years, this would correspond to $\dot M_{\rm emri} \sim 4\times10^{-5} M_\odot (\Sigma_{\rm add}/10^6$kg\,m$^{-2}$) of mass stripped per collision. To move substantially outside of the range already spanned by the disk, a substantial fraction of the EMRI mass would have to have been stripped. 

A mechanism for a disk instability in GSN 069 not (as far as we are aware) discussed in the literature thus far is the classical Hydrogen Ionization Instability (H.I.I), thought to be the origin of the outbursts in Cataclysmic Variables. The H.I.I occurs when different regions of an accretion flow have both ionized and neutral hydrogen, at which point the opacity in the disk becomes a strong function of disk temperature $\kappa \sim T_c^{10}$ \citep[e.g.,][]{Meyer1981}. Then the ability of the disk to radiatively cool begins to drop rapidly with central temperature $Q^{-} \sim T_c^4/\kappa \sim T_c^{-6}$, meaning that the hotter the disk becomes, the worse the disk gets at cooling. This is a classic setup for a runaway process. 

Hydrogen begins to become neutral when the disk temperature drops below $T\sim 10^4$ K, which readily occurs in the outer regions of our GSN 069 disk. In fact, roughly $\sim 40\%$ of the remaining mass of the disk is at these temperatures by $\sim$ late-2019 (Fig.~\ref{fig:2020_flare}), corresponding to $\sim 0.1 M_\odot$, interestingly close to the mass required to power the second flare. We suggest that the H.I.I may be an interesting avenue for future study, and may well be an interesting phenomena for TDE disks more generally.

\section{The Nature of `long-lived' TDEs}\label{sec:long-lived}

\begin{figure*}[]
	\centering
	\includegraphics[width=\textwidth]{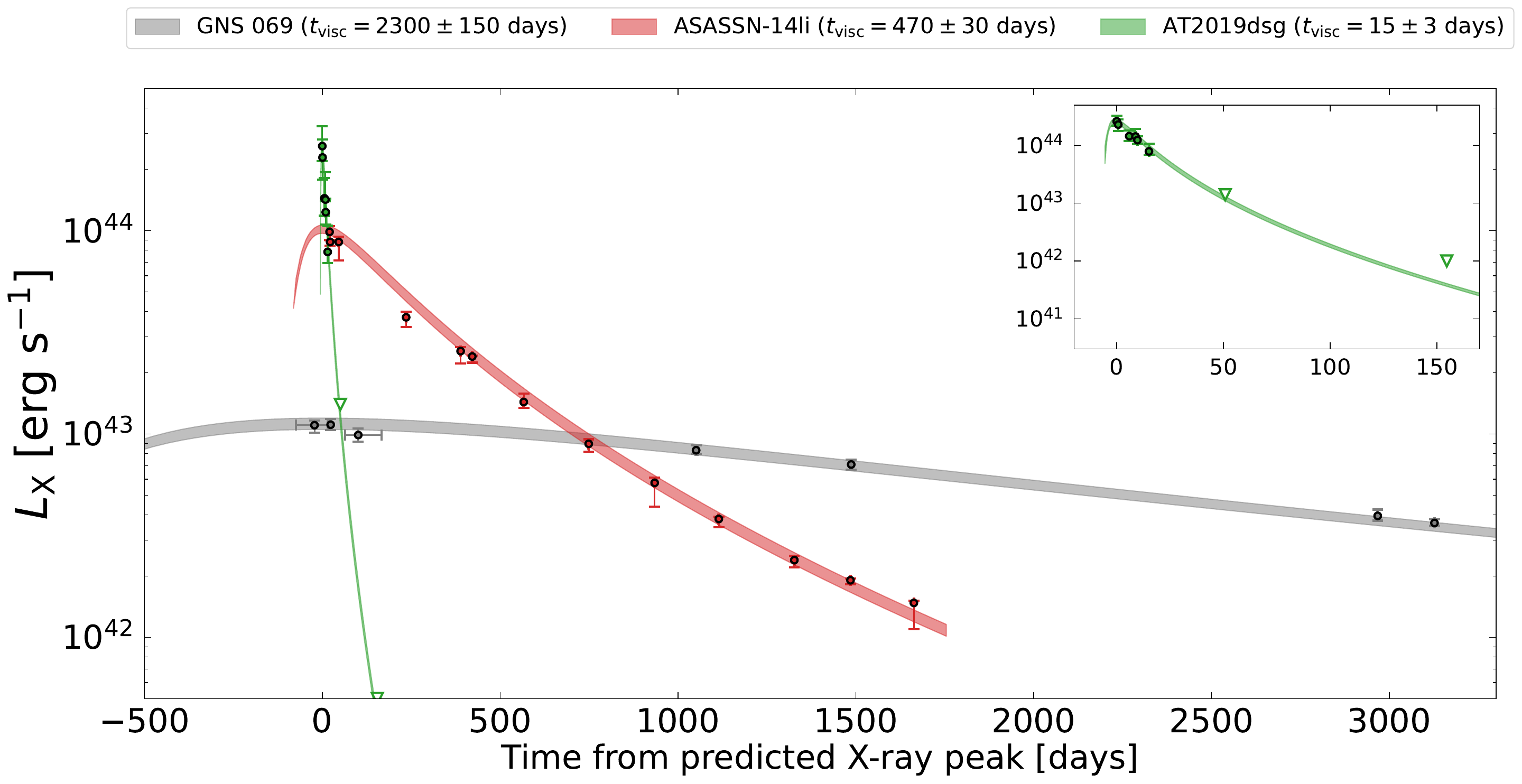}
 
	\caption{Comparison between 0.2-2.0 X-ray  luminosity light-curves between \target, ASASSN-14li and AT2019dsg. Shaded region represent the 68\% credible interval of \texttt{FitTeD-SPEC} fitting for all three sources. The figure lustrates how the same physical framework can describe the evolution of the three source, with $t_{\rm visc}$ been the main parameter driving their distinct time evolution. The main best-fitted parameters for the three sources is shown in Table \ref{tab:fit_comp}.}
    \label{fig:lc_compare}
\end{figure*}

\begin{figure*}
	\centering
    \includegraphics[width=0.8\textwidth]{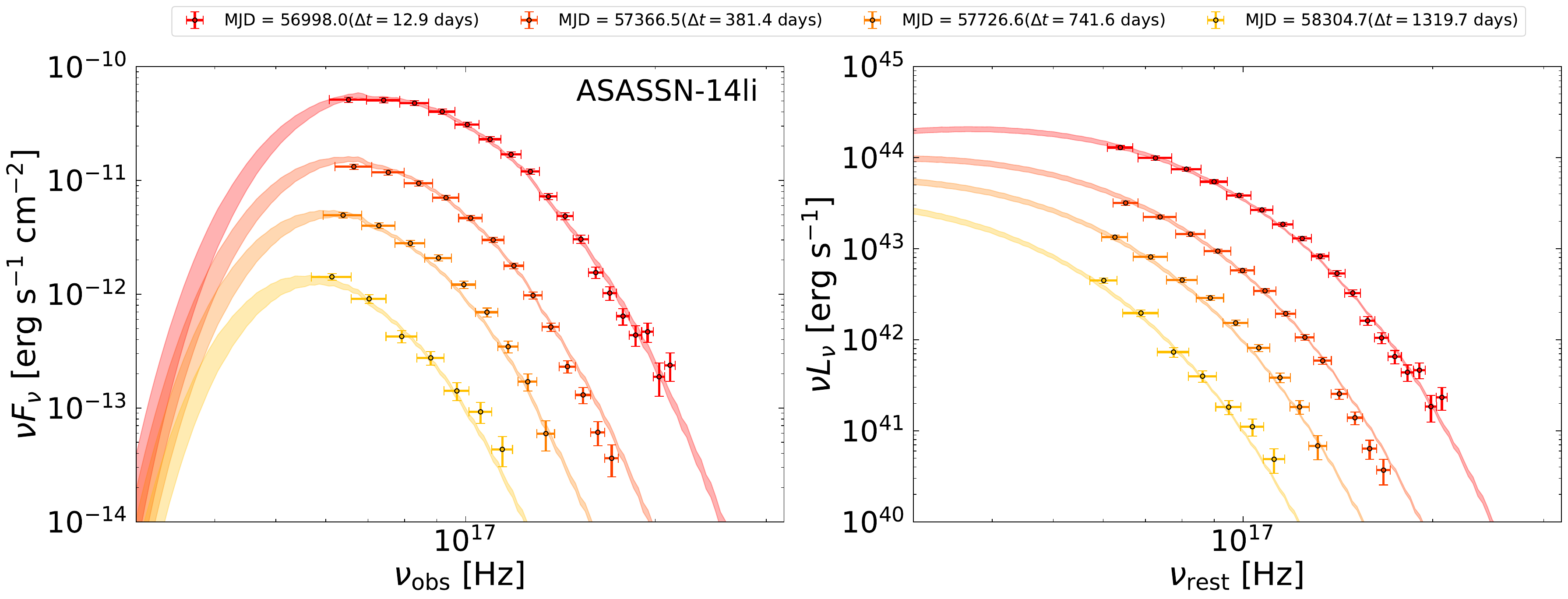}\\
    \vspace{0.5cm}
	\includegraphics[width=0.8\textwidth]{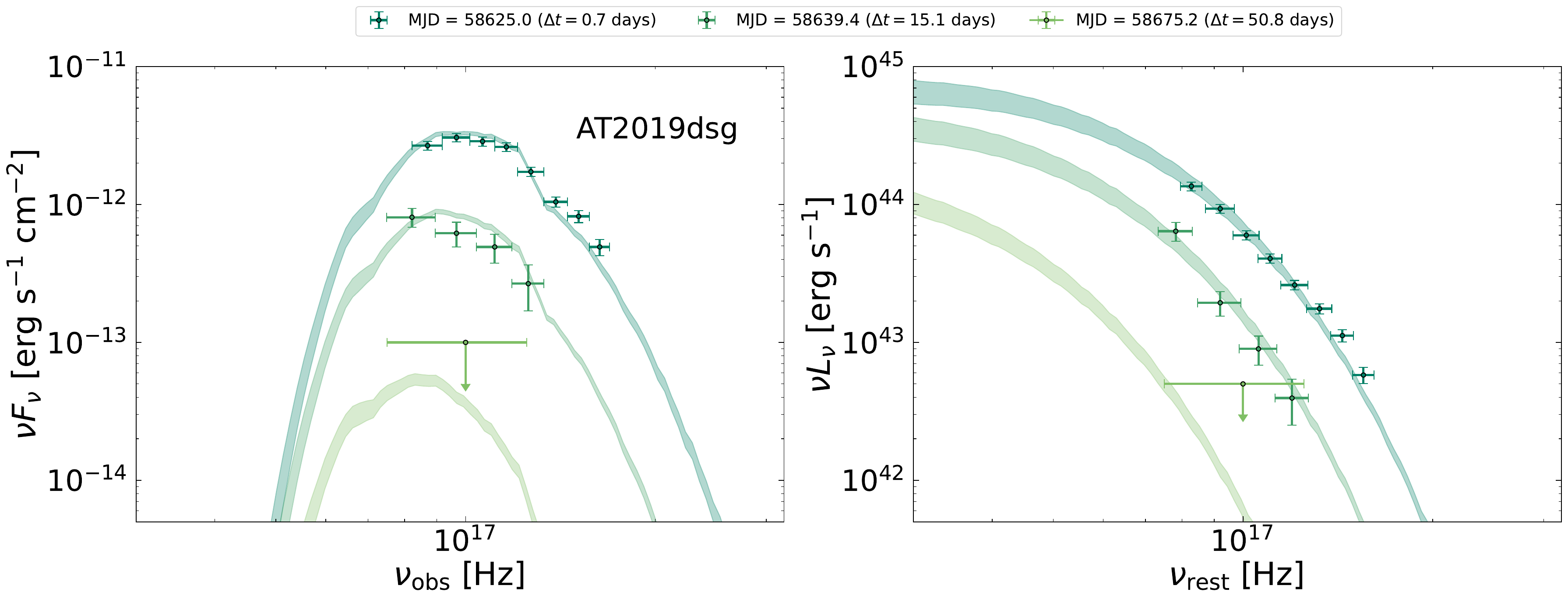}
 
	\caption{Examples of our time-dependent fitting on the X-ray spectral of ASASSN-14li (top) and AT2019dsg (bottom) at different epochs. Left: unfolded flux spectra with no absorption correction. Right: unfolded luminosity spectra, with both intrinsic and Galactic absorption corrections. The $\Delta t$'s in the legend, refer to the interval between the observation and the predicted X-ray peak.}
    \label{fig:14li_19dsg_SED}
\end{figure*}

\begin{figure}[]
	\centering
	\includegraphics[width=\columnwidth]{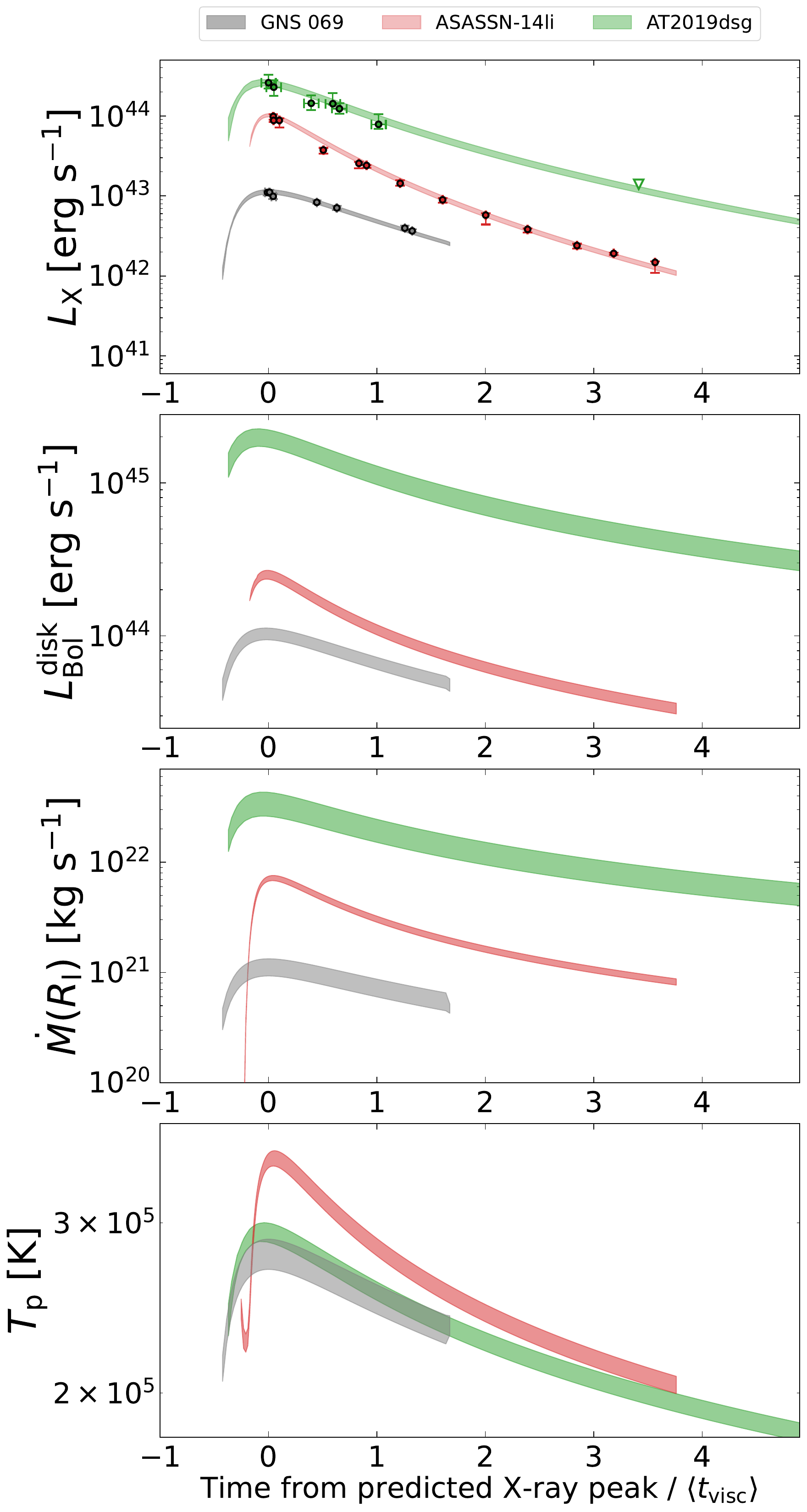}
 
	\caption{Evolution of observable and physical properties of \target, ASASSN-14li, and AT2019dsg, all normalized by the median value of the best-fit $t_{\rm visc}$ parameter (as listed in Table~\ref{tab:fit_comp}). This normalization demonstrates that the solutions to the disk evolution equations are scalable by $t_{\rm visc}$ and, to first order, follow the forms given in Equations~\ref{eq:Lbol_t}, \ref{eq:Md_t}, and \ref{eq:Tp_t}. The remaining differences in their properties—primarily in the amplitudes—are governed by the other disk-back hole parameters, while the time evolution becomes largely universal once normalized by $t_{\rm visc}$.}
    \label{fig:lc_compare_norm}
\end{figure}

\begin{figure}[]
	\centering
	\includegraphics[width=\columnwidth]{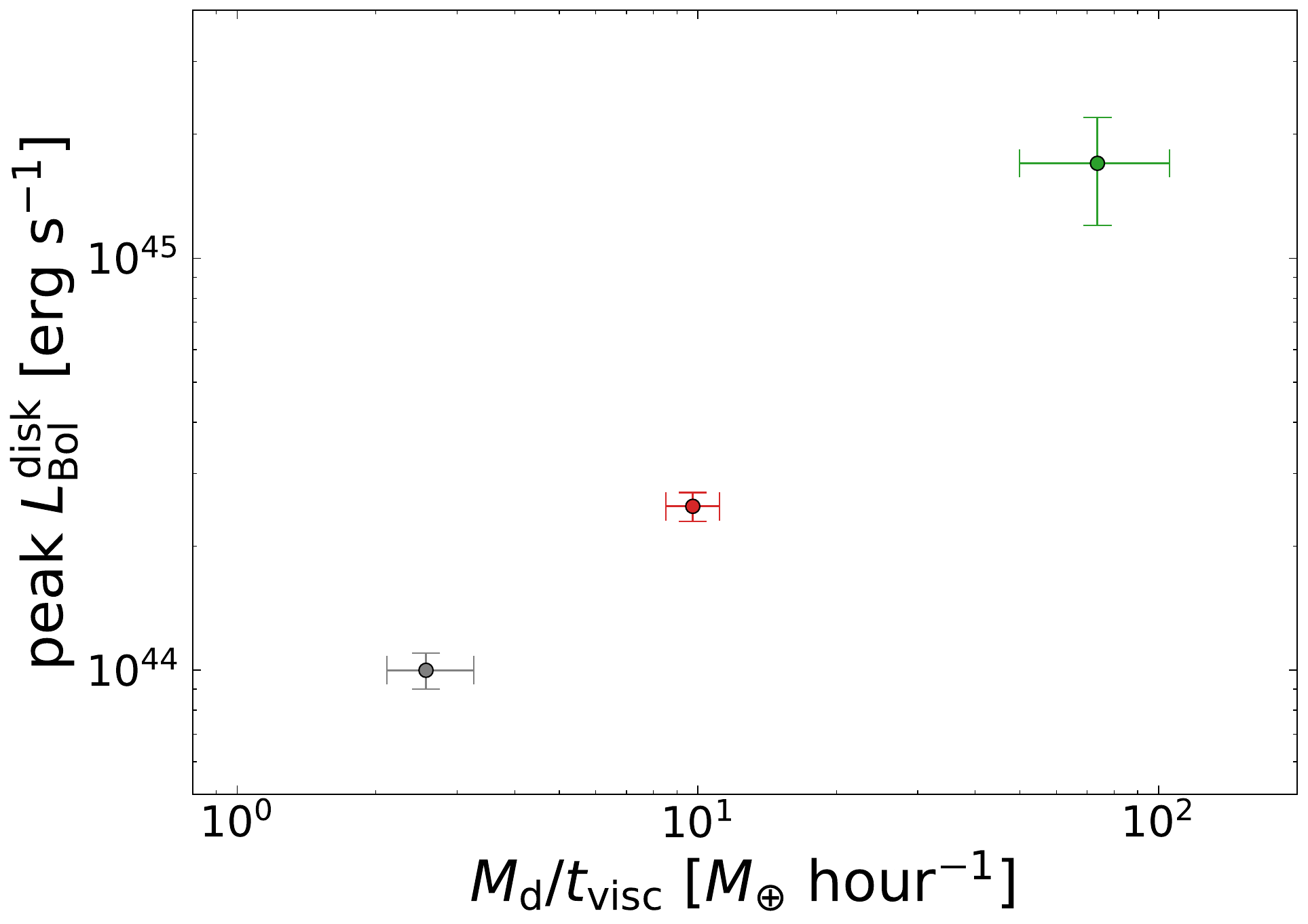}
 
	\caption{Correlation between peak disk bolometric luminosity of \target (grey), ASASSN-14li (red) and AT2019dsg (green), and the ratio between their best-fitted initial disk mass ($M_{d}$) and `viscous' timescale ($t_{\rm visc}$). The figure illustrate how even disks with lower mass content (AT2019dsg), will produce brighter peak luminosities, if the rate of angular momentum transport (i.e., $W^r_{\ \phi} \propto t_{\rm visc}^{-1}$), is higher, such that the absolute mass accretion rate, $\dot{M}(r_I)$, is higher.}
    \label{fig:compare_Lbol_Md_tv}
\end{figure}

It is quite clear that \target\ X-ray emission evolves slowly—very slowly—particularly when compared to more ‘typical’ TDEs. Perhaps even more physically insightful than its slowly declining X-ray luminosity is the related, behavior of its inferred `inner' disk temperature. 

In this section, we compare \target\ to TDEs that exhibit more `typical' X-ray emission, such as ASASSN-14li and AT2019dsg.
Even before applying our time-dependent relativistic disk modeling, the contrast in color-temperature evolution ($T_{\rm BB,X}$)\footnote{We refer to the color-temperature or apparent temperature ($T_{\rm BB,X}$) as the parameter derived from \texttt{XSPEC} models that are either not disk models (e.g., \texttt{blackbody}), or those which lack sufficient disk physics (e.g., \texttt{diskbb}), such as color corrections or consistent boundary conditions \citep[see discussion in][]{Zimmerman2005,Mummery_2021,Guolo_Mummery2025}. This is such that their derived $T_{\rm BB,X}$ can not be interpreted as, and will be always higher than, the peak physical temperature ($T_p$) of a more realistic disk.} among these sources is striking, as already shown by previous studies. Over the span of $\sim3300$ days (from 2010 to 2019), the color-temperature of \target\ declined by only $\sim$25\% \citep{Miniutti2023_longterm}. For ASASSN-14li, a $\sim$45\% decline occurred over $\sim1600$ days (2014 to mid-2019) \citep{Ajay2024}, while AT2019dsg exhibited a similar 45\% decline in just 15 days \citep{Cannizzaro2021}. This leads to dramatically different average disk cooling rates: $\mathrm{d}T_{\mathrm{BB,X}}/\mathrm{d}t \approx -50$ K day$^{-1}$ for \target, $-220$ K day$^{-1}$ for ASASSN-14li, and $-15,000$ K day$^{-1}$ for AT2019dsg. Nevertheless, the functional form of their cooling rate can be well describe by disk theory \citep{Guolo2024}, i.e., by Eq. \ref{eq:Tp_t}.

For the cases of ASASSN-14li and AT2019dsg, it has already been shown \citep{Mummery2020, Mummery2024fitted} through light curve fitting with \texttt{FitTeD}, that their emission---X-ray at all times and the late-time UV/optical plateau---can be described by an evolving relativistic accretion flow, i.e., the framework introduced in \S\ref{sec:theory}. Having now extended this framework to include X-ray spectral fitting, we applied \texttt{FitTeD-XSPEC} to the available data for these sources: \xmm\ observations of ASASSN-14li and \nicer\ observations for AT2019dsg \citep[using data reduction as described following, respectively,][]{Ajay2024, Cannizzaro2021}, as detailed in Appendix~\ref{app:data}.

Given that the physics driving the evolution of the multi-wavelength emission ASASSN-14li and AT2019dsg has been explored in previous work \citep{Mummery2020, Mummery2021b, Mummery2024fitted}, our emphasis here is on their X-ray spectral evolution as revealed by new time-dependent spectral fitting using \texttt{FitTeD-XSPEC}. 
A comparison of the X-ray luminosity evolution for \target, ASASSN-14li, and AT2019dsg---all modeled within the same framework---is shown in Fig.~\ref{fig:lc_compare}, and the best-fit values for the most relevant model parameters are listed in Table~\ref{tab:fit_comp}. For completeness, in Fig.~\ref{fig:14li_19dsg_SED} we show that our modeling not only reproduces the integrated X-ray luminosity evolution of ASASSN-14li and AT2019dsg, but also captures their spectral evolution—specifically, the cooling rate as a function of time.

While Fig.~\ref{fig:lc_compare} highlights the wide range of timescales over which TDEs can evolve in the X-ray band, it also demonstrates the flexibility of the time-dependent disk model (\S\ref{sec:theory}), as the emission from three markedly different sources can be described within the same framework. From Table~\ref{tab:fit_comp}, we find that although the key model parameters differ among the sources, the viscous timescale stands out as the most strikingly distinct: AT2019dsg and ASASSN-14li exhibit \( t_{\rm visc} = 15 \pm 3 \) days and \( t_{\rm visc} = 470 \pm 30 \) days, respectively, while for \target, we find \( t_{\rm visc} = 2300 \pm 150 \) days. This confirms, unsurprisingly, that the viscous timescale is the primary parameter controlling the disk's temporal evolution. Indeed, once the light curves are rescaled by their best-fit viscous timescales, as shown in Fig.~\ref{fig:lc_compare_norm}, the apparent differences in their time evolution largely disappear. To first order, the evolution of their disk properties follows the analytical scalings described by Equations~\ref{eq:Lbol_t}, \ref{eq:Md_t} and \ref{eq:Tp_t}.

\begin{deluxetable}{C|CCC}
\tablenum{2}
\label{tab:fit_comp}
\tabletypesize{\small}
\tablecaption{Summary of inferred parameters for \texttt{FitTeD-XSPEC} fitting of \target, ASASSN-14li and AT2019dsg.}
\tablehead{ \colhead{\rm Parameter} &  \colhead{\target} &  \colhead{ASASSN-14li}& \colhead{AT2019dsg} }
\startdata
{\rm log} \  (M_{\bullet}/M_{\odot}) & 6.55 \pm 0.15 & 6.35 \pm 0.10 & 6.80 \pm 0.15 \\ 
M_{d} \ [M_{\odot}] &  0.40^{+0.15}_{-0.10} & 0.30^{+0.05}_{-0.03} &0.08^{+0.05}_{-0.03} \\
$t_{\rm visc}$ \  [{\rm days}] &  2300 \pm 150 & 470\pm 30 &15 \pm 3 \\
{\rm peak}( L_{\rm Bol}^{\rm disk}) \  [{\rm 10^{44} \  erg \ s$^{-1}$}] & 1.0^{+0.2}_{-0.1}  & 2.5_{-0.3}^{+0.1} &  18^{+5}_{-6}
\enddata 
\tablecomments{Central values represent the medians of the posterior with uncertainties corresponding to their 68\% credible intervals.}
\end{deluxetable}

The important questions are then how to interpret this, why distinct sources have distinct values of $t_{\rm visc}$, and what this means physically. To interpret the viscous timescale it is illustrative to move to the $\alpha$ model framework of \cite{Shakura1973}, while bearing in mind the limitations of employing such a simple description of disk turbulence. What is required is to estimate the scale of $W^r_\phi$, which sets the evolutionary timescale of these systems. The stress $W^r_\phi$ is defined as being equal to the correlations in the turbulent disk velocity \citep{Balbus1999}
\begin{equation}
    W^r_\phi \equiv r \left\langle \delta U^r \delta U^\phi\right\rangle .
\end{equation}
The $\alpha$ model then simply assumes these velocity scales will be $\sim$ the speed of sound, an argument effectively based on dimensional analysis. Then 
\begin{equation}
    \delta U^r \sim c_S, \quad \delta U^\phi \sim  c_S, \quad W^r_\phi \sim r c_S^2 = \alpha r c_S^2 ,
\end{equation}
where $c_S$ is the sound speed, and $\alpha$ is a dimensionless number which is {\it a priori} unknown, but is expected to be $\alpha\leq 1$ so that the turbulence does not cause shocks within the disk. Note that the above description is equivalent to our $\mu=0$ parameterisation, if one assumes that the disk has a constant aspect ratio $H/R = {\rm constant}$. This is because the solution of vertical hydrostatic equilibrium in the disk
\begin{equation}
    {\partial P \over \partial z} \approx - {GM_\bullet \rho z\over r^3}  \to  c_S^2 \approx \left({H\over R}\right)^2 v_K^2 ,
\end{equation}
implies a constant stress 
\begin{equation}
    W^r_\phi = \alpha (H/R)^2 GM_\bullet = {\rm constant}, 
\end{equation} 
for fixed aspect ratio. Then, taking $w = \alpha (H/R)^2 GM_\bullet$, and substituting into the definition of the viscous timescale (eq. \ref{eq:t_visc}), one finds the classical result 
\begin{equation}
    t_{\rm visc} = {2\over 9} \alpha^{-1} (H/R)^{-2} \sqrt{r_0^3\over GM_\bullet} .
\end{equation}
Taking $r_0$ equal to the tidal radius divided by the impact parameter $\beta$, we find
\begin{equation}
    t_{\rm visc, TDE} \propto \alpha^{-1} (H/R)^{-2} \sqrt{R_\star^3\over \beta^3 GM_\star} .
\end{equation}
The question is then, can the above quantity vary by 2 orders of magnitude across a population of TDEs (i.e., the range seen in Table \ref{tab:fit_comp})? To which the answer is clearly yes. Taking a standard mass-radius relationship for a main sequence star $R_\star \propto M_\star^{4/5}$, one finds 
\begin{equation}
    t_{\rm visc, TDE} \propto \alpha^{-1} (H/R)^{-2} \beta^{-3/2} M_\star^{7/10} .
\end{equation}
TDE stellar masses will vary by at least $\sim 1$ order of magnitude, as can $\beta$, which already could explain $\sim 2$ orders of magnitude variance in $t_{\rm visc}$. Allowing for even moderate changes in $\alpha$ or $H/R$ would then comfortably explain the observed variance.

Even if all TDEs were formed from stars of the same mass being disrupted on $\beta =1$ orbits, the resulting disks will exhibit distinct viscous timescales, \(t_{\rm visc}\), owing to the intrinsically variability in \(\alpha\)-type parameters which will be produced in a population of disks (and of course the variability in $H/R$). It is important to emphasize that the viscous timescale, \( t_{\rm visc} \) (or $\W$, or $\alpha$-type parameters) should not be assumed to be universal or constant across the population of newly formed accretion disks. Instead, they should be treated as free parameters that can be constrained observationally, as it sets the time evolution of many of the system observables. The theoretical framework developed in \citet{Mummery2020}, implemented as a light curve fitting tool \citep[\texttt{FitTeD};][]{Mummery2024fitted} and here extended to time-dependent X-ray spectral fitting (\texttt{FitTeD-XSPEC}), provides a robust method to do so, and is of broad theoretical interest.

Of course, although the time-scale of the X-ray emission is mostly controlled by $t_{\rm visc}$, other parameters also effect the emission. For example, although AT2019dsg has the lowest mass content in its disk, it has the brightest peak X-ray and peak Bolometric disk luminosity ($L_{\rm Bol}^{\rm disk}$), this occurs because the peak the disk emission $L_{\rm Bol}^{\rm disk}$ is primarily dependent on the peak rate at which the matter flows trough the inner disk region, i.e., the absolute accretion rate ($\dot{M}(r_I)$) near the ISCO (see third panel in Fig.\ref{fig:lc_compare_norm}), which is a complicated non linear function on the many free parameters of the theory, but correlates at first order with the ratio $\sim M_{\rm d}/t_{\rm visc}$, as we show in Fig.~\ref{fig:compare_Lbol_Md_tv}.

This again highlights the crucial role of $t_{\rm visc}$ in the system’s evolution: a low $t_{\rm visc}$, or equivalently a high $W^r_\phi$ (i.e., efficient angular momentum transport), leads to rapid diffusion of disk material (Eq.~\ref{eq:fund}). As a result, even a low-mass disk can produce a high peak accretion rate—and thus high luminosity—because the mass flows inward more quickly, though the X-ray emission will be shorter lived, given the inner disk mass will be rapidly accreted. The opposite is true for \target, whose high $t_{\rm visc}$ results in slower diffusion and correspondingly lower peak accretion rates.

It is important to emphasise, while the Bolometric disk luminosity of TDE disks decays as a power-law (with index depended on some details of the physics of the inner disk conditions Eq. \ref{eq:Lbol_t}), contrary to what is often assumed, the X-ray light curves will not follow a power-law. This would be true {\it even if} the accretion rate was equal to the fall-back rate. Instead the X-ray light curve evolves with the functional form of a power-law temporal decay times an exponential dependence on the cooling peak disk temperature. See detailed discussion in \citet{Mummery2020}.

From this, we conclude that the slow evolution of \target\ arises naturally within the framework of time-dependent accretion disk theory, without requiring any modification to the underlying physics—only moderate changes to free parameters, all well within plausible bounds expected from TDE systems. 

It seems natural that the same explanation can be extended to account for the decade-long evolution observed in all `long-lived' TDEs, such as: 2XMM J123103.2+110648 \citep{Terashima2012,Ho2012,Lin2017b}, 2XMMi J184725.1-631724 \citep{Lin2011,Lin2018}, 3XMM J150052.0+015452 \citep{Lin2017b,Lin2022} and 3XMM J215022.4-055108 \citep{Lin2018b,Lin2020}. 
It also seems natural, that slowly evolving X-ray TDEs are more likely to be discovered via X-ray surveys, as is the case for all the sources mentioned above. 

One may ask whether even more extreme cases of slow evolution in soft, thermal X-ray sources can also be explained within this same theoretical framework, simply by adopting even longer viscous timescales. For example, RX~J1301+27 \citep[][also a QPE source]{Sun2013,Giustini2020,Giustini2024} has shown only a factor of $\sim 2$ change in X-ray luminosity over the $\sim 30$ years since its discovery. Similarly, the QPE source eRO-QPE2 \citep{Arcodia2021} had shown no detectable X-ray emission in the mid-1990s by \rosat, but has exhibited effectively no evolution in its 'quiescent' disk emission since its discovery in 2020—when it was identified both as an X-ray source and a QPE emitter \citep{Arcodia2024ero2}. It certainly does not seem unreasonable that $t_{\rm visc}$ could vary by $\sim 3$ orders of magnitude across a TDE population, given its strong dependence on 4 parameters all of which can vary substantially. This is a possibility that can be tested from the data using the tools put forward in this work.

Many of the sources discussed above have often been referred to as `super-soft AGN'. However, this AGN classification is primarily based on their location above the \citet{Kewley2001} line in BPT-like diagnostic diagrams, inferred from narrow emission line ratios, and by the fact they appear to be longer-lived, e.g., as compared to more naive TDE fall-back rate assumptions as discussed previously. We now know, however, from both observations \citep{Short2023,Newsome2024} and \texttt{CLOUDY} simulations \citep{Patra2023, Mummery2025EELR}, that TDEs can themselves produce compact narrow-line regions (NLRs) in their host galaxies. If host nuclear optical spectrum is taken sufficiently long after the disk formation, these NLRs can place the host in the `AGN region' (in reality, simple a region where the ionization source is hotter than massive stars) of diagnostic diagrams—even in the absence of any prior nuclear activity. See \citet{Mummery2025EELR} for a detailed discussion.

The findings presented here—namely, that \target's X-ray emission is naturally and self-consistently reproduced by the same physical framework used to model `typical' TDEs—call into question the existence itself of corona-less `super-soft AGN'. Rather, our results suggest that these sources may simply be TDEs, sampling the broad range of viscous timescales that newly formed accretion disks can realize in nature. The notion of an AGN lacking a hard X-ray corona, exhibiting ultra-soft X-ray spectra, and showing no persistent broad-line region, or dusty torus, may in fact, be an observational mirage—arising from the misclassifications of slowly evolving TDEs.

\section{Conclusion}\label{sec:conclusion}

In this paper, we have explored the properties of the `long-lived' tidal disruption event (TDE) and first quasi-periodic eruption (QPE) source, \target, within the framework of relativistic time-dependent accretion disk theory (\S\ref{sec:theory} and references therein). Our key accomplishments and results are as follows:

\begin{itemize}
    \item We implemented the light curve fitting package \texttt{FitTeD} \citep{Mummery2024fitted}—which solves the relativistic time-dependent accretion disk equations and computes the evolving spectral energy distribution (SED) as a function of time—into the standard X-ray fitting package \texttt{pyXspec} (\S\ref{sec:fitted_xspec} and Fig.~\ref{fig:fitted_scheme}). Allowing simultaneous and self-consistent fitting of light-curve and spectral data.
    
    \item We used \texttt{FitTeD-XSPEC} to fit seven epochs of X-ray spectra and two epochs of UV spectra from \target, spanning from its discovery in 2010 through late 2019 (Fig.~\ref{fig:lc+SED}).
    
    \item We recovered the time evolution of several disk and black hole properties—some for the first time, such as the surface density profile $\Sigma(r, t)$, and others with unprecedented precision (Figs.~\ref{fig:corner} and ~\ref{fig:profiles}). We also discussed the limitations of the current approach (\S\ref{sec:limitations}).
    
    \item We examined the implications of our results in the context of QPE models. In agreement with our previous time-independent SED modeling \citepalias{Guolo2025}, we found that QPEs are present at low accretion rates and absent at high accretion rates (Fig.~\ref{fig:kaur}). This trend is the \textit{opposite} of what is predicted by currently available disk instability models for QPEs.
    
    \item In the context of orbiter-disk collision models, we confirmed the disk had already substantially expanded in 2014—when no QPEs were present—in agreement with \citetalias{Guolo2025}. However, with our novel constraints on $\Sigma(r, t)$, we find that the specific scenario proposed by \citet{Linial2023} --in which the absence of QPEs in 2014 and their emergence in 2018 result from the lower temperature of the QPEs in 2014 -- appears difficult to reconcile with the observed evolution in $\Sigma(r_{\rm orb}, t)$, which is smaller than what would be required for such a transition (Figs.~\ref{fig:sigma_hist} and \ref{fig:LM23}).
    
    \item We argue that the absence of QPEs in 2014 and their appearance in 2018 in \target\ cannot be easily explained by any currently available model, addressing this problem should be a central focus of future theoretical efforts. Particularly, given that the properties of \target's disk at these epochs is well understood.

    \item In a more speculative way, we discuss the origin of the 2020–2022 rebrightening event observed in \target (\S\ref{sec:2020}), evaluating the feasibility of previously proposed scenarios within the framework of time-dependent disk theory. We also put forward an alternative explanation—the Hydrogen ionization instability—a phenomenon well known in stellar-mass accretion systems such as Cataclysmic Variables. While promising, this possibility remains preliminary and warrants further investigation.
    
    \item We discussed the `long-lived' nature of \target\ by comparing our results with \texttt{FitTeD-XSPEC} fits of two `typical' TDEs, ASASSN-14li and AT2019dsg. We showed that the X-ray emission and evolution of all three sources can be described by the same model (Figs.~\ref{fig:lc_compare} and \ref{fig:lc_compare_norm}), with the primary difference being the `viscous' timescale ($t_{\rm visc}$), which varies by two orders of magnitude among them.

    \item We discuss the role of $t_{\rm visc}$ in a realistic, time-dependent view of TDE accretion disks (\S\ref{sec:long-lived}), and how it is mostly likely the key to understand the nature of all the so-called long-lived TDEs (a.k.a, super-soft ``AGN"), as well the the distinct evolution time-scales in distinct TDEs. 

\end{itemize}

\textit{Acknowledgements} -- 
MG is supported in part by NASA XMM-Newton grant 80NSSC24K1885.
This work was supported by a Leverhulme Trust International Professorship grant [number LIP-202-014]. MN is supported by the European Research Council (ERC) under the European Union’s Horizon 2020 research and innovation programme (grant agreement No.~948381) and by UK Space Agency Grant No.~ST/Y000692/1. AI acknowledges support by the Royal Society. The HST data presented in this article were obtained from the Mikulski Archive for Space Telescopes (MAST) at the Space Telescope Science Institute. The specific observations analyzed can be accessed via \dataset[doi:10.17909/xn57-rw37]{http://dx.doi.org/10.17909/xn57-rw37}. STScI is operated by the Association of Universities for Research in Astronomy, Inc. under NASA contract NAS 5-26555.

\bibliography{tde}{}
\bibliographystyle{aasjournal}

\appendix
\section{Supplementary Mathematical Formalism}\label{app:math}
We provide here the relevant relations required to complete the physical framework presented in \S\ref{sec:theory}. The functional forms of the four-velocity components for circular orbits in the equatorial plane of the Kerr metric:

\begin{align}
U_0 &= -{{\cal A} \over {\cal D}^{1/2}}, \quad 
  U^0 = {{\cal B} \over {\cal D}^{1/2}}, \quad 
 U_\phi = \sqrt{GM_\bullet r} {{\cal C} \over {\cal D}^{1/2}} , \nonumber \\ 
 &\quad U^\phi = \sqrt{{GM_\bullet}\over {r^3}} \, {1\over {\cal D}^{1/2}} , \quad  
  U_\phi ' ={1\over 2}\sqrt{GM_\bullet\over r} {{\cal BE} \over {\cal D}^{3/2}} , \nonumber \\  
 &\quad\quad\quad\quad\quad \Omega \equiv \frac{U^\phi}{U^0} = \sqrt{{GM_\bullet }\over {r^3}} {1\over {\cal B}} ,  \label{orb_eqs} 
 \end{align}
 where ${\cal A, B, C, D}$ \& ${\cal E}$ are all relativistic corrections which tend to 1 at large radii \citep[cf.][]{Novikov1973}
 \begin{align}
{\cal A} &= 1-2r_g/r +a\sqrt{r_g/r^3}, \\
{\cal B} &= 1+a\sqrt{{r_g}/{r^3}}, \\
{\cal C} &= 1 + {a^2}/{r^2} - 2a\sqrt{{r_g}/{r^3}}, \\
{\cal D} &=  1 - {3r_g}/{r} + 2a\sqrt{{r_g}/{r^3}}, \\
{\cal E} &= 1 - 6r_g/r - 3a^2/r^2 + 8a\sqrt{r_g/ r^3} . 
\end{align}
Note that ${\cal E}(r_I)=0$ defines the ISCO radius, the inner limit of validity of our model.

The Green function solution for the disk fluid diffusion equations (Eq. \ref{eq:fund}), for all times $t > t_0$ (and radii $r>r_I$) take the functional form 
\begin{multline}\label{green_s}
G_\Sigma(r, t; r_0, t_0) = {M_d \over 2\pi r_I^2 c_0}  \sqrt{x^{-\alpha} f_\alpha(x) \exp\left(-{1 \over x} \right) \left(1 - {2\over x}\right)^{5/2 - 3/4\alpha} }\\ 
 {x^{-3/4 - \mu} \over \tau} \exp\left({-f_\alpha(x)^2 - f_\alpha(x_0)^2 \over  4\tau} \right) I_{1\over 4\alpha} \left({ f_\alpha(x) f_\alpha(x_0) \over  2\tau}\right),
\end{multline}
where 
\begin{equation}
c_0 = x_0^{(1 + 14\mu)/8} \left(1 - {2\over x_0}\right)^{3/4 - 3/8\alpha} \left( f_\alpha(x_0)\exp\left({1\over x_0}\right) \right)^{-1/2} ,
\end{equation}
is a normalisation factor ensuring that $M_d$ is the total initial disk mass. In this expression, $I_\nu$ is the modified Bessel function of order $\nu$, and $\alpha$ is related to the stress index $\mu$ via $\alpha = (3-2\mu)/4$. The function $f_\alpha(x)$ is given by 
\begin{equation}
    f_\alpha(x) = {x^\alpha \over 2 \alpha} \sqrt{1 - {2\over x}}\left[1  - {x^{ - 1} \over { (\alpha - 1)}} {}_2F_1\left(1, {3\over 2}-\alpha; 2-\alpha; {2\over x}\right) \right] \\ + {2^{\alpha - 2} \over \alpha (\alpha - 1)}\sqrt{\pi} {\Gamma(2-\alpha)  \over  \Gamma({3/ 2} - \alpha)} ,
\end{equation}

where $_2F_1(a, b; c; z)$ is the hypergeometric function, $\Gamma(z)$ is the gamma function, and 
\begin{equation}
x \equiv 2 r / r_I. 
\end{equation}
The variable $x_0 = 2r_0/r_I$ is the (normalised) initial location of the disk material. The time variable $\tau$ is given by 
\begin{equation}
\tau \equiv \sqrt{2 \over GMr_I} {w \over r_I} \left(1 - {r_I \over r_0}\right) \left(t -t _0\right),
\end{equation} 
where $t$ is measured in physical units. It is important to note that $\tau$ as defined here is {\it not} equal to the time in units of the viscous timescale at the initial radius $\tau \neq (t-t_0)/t_{\rm visc}(r_0)$
\section{Data Reduction and Treatment}\label{app:data}
\subsection{\target X-ray data reduction and processing}
The X-ray data set underlying this work is based on \xmm and \swift/XRT observations. 
The \xmm observations  were taken as part of Guest Observer (GO) and Director discretionary Time (DDT) programs (PIs Saxton and Miniutti, OBS-ID 0740960101,0657820101, 0823680101 and 0851180401) and are publicly available in the \xmm archive. The observations were taken in Full Frame mode with the thin filter using the European Photon Imaging Camera \citep[EPIC;][]{Struder2001}. The observation data files (ODFs) were reduced using the \xmm Standard Analysis Software \citep[SAS;][]{Gabriel_04}.
The raw data files were then processed using the \texttt{epproc} task. 
Since the pn instrument generally has better sensitivity than MOS1 and MOS2, we only analyze the pn data. 
Following the \xmm data analysis guide, to check for background activity and generate ``good time intervals'' (GTIs), we manually inspected the background light curves in the 10--12\,keV band. 
Using the \texttt{evselect} task, we only retained patterns that correspond to single and double events (\texttt{PATTERN<=4}). The source spectra were extracted using a source region of $r_{\rm src} = 35^{\prime\prime}$ around the peak of the emission. 
The background spectra were extracted from a $r_{\rm bkg} =
108^{\prime\prime}$ region located in the same CCD. The ARFs and RMF files
were created using the \texttt{arfgen} and \texttt{rmfgen} tasks,
respectively. For the 2018 and 2019 observations, in which QPE are present we select only GTIs where QPEs are not present, as similarly as illustrated in Fig.~10 of \citetalias{Guolo2025}, such that the `quiescent' disk emission is the resulting spectrum; no QPEs were present in the 2010 or 2014 \xmm data.

The source was observed by the \swiflong\ X-ray Telescope (XRT) on several occasions. Most individual observations are too short to yield useful spectral information, so we stacked groups of consecutive observations using an automated online tool\footnote{\url{https://www.swift.ac.uk/user_objects}} \citep{Evans2009}. This was done for three distinct periods, as detailed in Table~\ref{tab:data}. Stacking not only improves the signal-to-noise ratio---allowing for a fittable spectrum with sufficient counts---but also averages out short-term variability that cannot be captured by the analytical framework described in \S\ref{sec:theory}.

\begin{figure}[h!]
	\centering
	\includegraphics[width=\columnwidth]{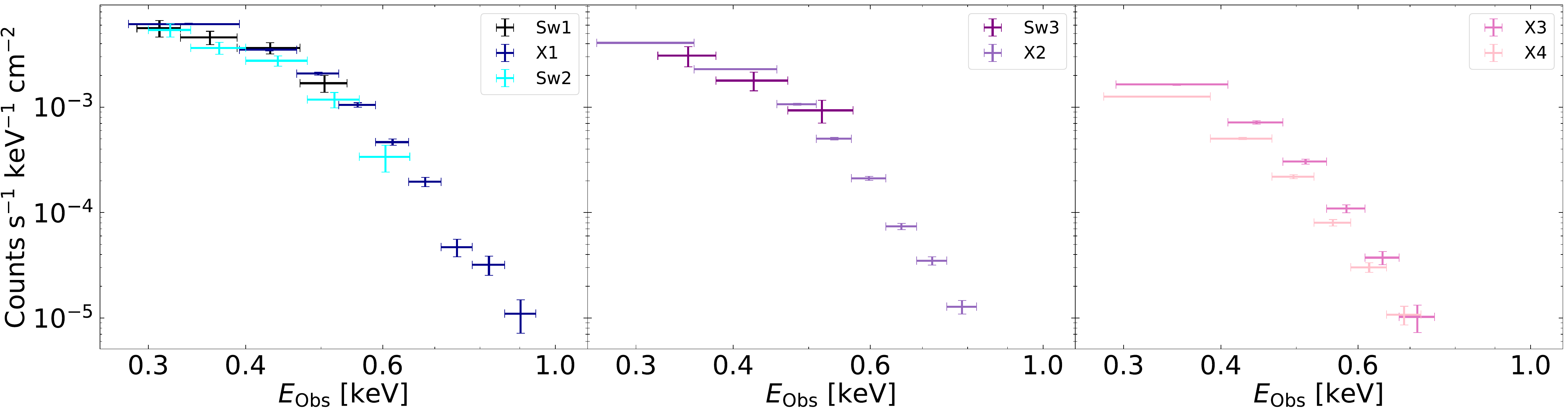}
 
	\caption{The seven folded X-ray spectra normalized by their instrument's effective area as function of energy. }
    \label{fig:x_reduction}
\end{figure}

The seven final X-ray spectra (four from \xmm\ and three from \swift/XRT), each normalized by the effective area of the corresponding instrument (i.e. simple setting,\textit{`Plot.area = True'} in \texttt{pyXspec}), are shown in Fig.~\ref{fig:x_reduction}. We note, as will be relevant in the main section, that visually, spectra Sw1, X1, and Sw2 appear to have consistent effective area-normalized count rates and similar spectral shapes, within uncertainties.

\subsection{\target UV data processing}
\begin{figure}[h!]
	\centering
	\includegraphics[width=\columnwidth]{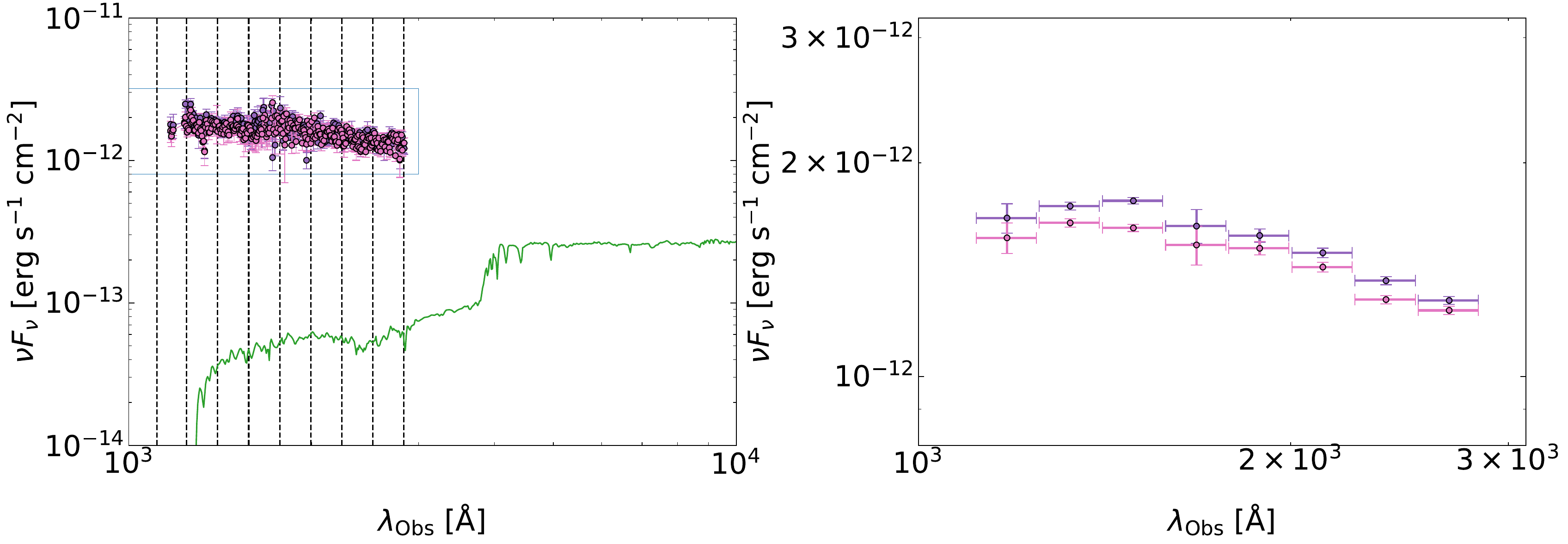}
 
	\caption{UV spectra processing. Right: observed unbinned spectra in the two epochs 2014 (purple) and 2018 (pink). Green spectra shows best-fitted stellar contribution modeled in \citetalias{Guolo2025}. Dotted lines show the wavelength range used to bin the spectra. Left: binned spectra. All the spectra are shown as observed, with no extinction correction. }
    \label{fig:uv_reduction}
\end{figure}

The reduction \hst data was already discussion in \citetalias{Guolo2025} which we simple follow. In this work, we have however processed the data before the multi-wavelength fitting, first we have subtracted the underline stellar population (which was simultaneous fitted in \citetalias{Guolo2025}), the stellar population accounts for $\leq 5\%$ of the observed flux at the entire available wavelength range (3000-1150 \AA). The stellar component removal was done by simple subtracting the best-fitted stellar model (Green component in the left panel of Figure 4 in \citetalias{Guolo2025}) from the two observed spectra. Finally, we have binned the spectra, from its original sampling to a series of eight log-space synthetic narrow band filters (as illustrate in Fig.~\ref{fig:uv_reduction}). This binning makes the number of model computations substantially decrease, while preserving all the spectral information, e.g., wavelength range and spectral shape, given the spectra are continuum dominated. The only mildly bright emission lines, e.g. Ly$\alpha$, were masked before the binning. Some basic properties of the spectra are discussed in the main section, while a detail description is presented in \citetalias{Guolo2025}.

\subsection{ASASSN-14li and AT2019dsg data}

All available \xmm\ spectra of ASASSN-14li were used in this work. The data reduction, including corrections for pile-up, is described in \citet{Ajay2024}; we simply use the same processed data. For AT2019dsg, we used all X-ray spectra with detections above the background level from \nicer, as presented in \citet{Cannizzaro2021}. We follow the exact same procedure, producing six spectra from observations 200680101–200680108, using the same reduction and stacking steps as described in \citet{Cannizzaro2021}. We also use two late-time upper limits from \swift/XRT and \xmm as presented in \citet{Guolo2024}. We do not use early time \swift/XRT data as no fittable spectra can be obtained from their low count rate. Optical and UV data for both sources were obtained from the \textit{ManyTDE} database\footnote{\url{https://github.com/sjoertvv/manyTDE}}, whose reduction methods are described in \citet{Mummery2024}.

\section{Constraining the maximum $T_{\lowercase{obs}}^{\lowercase{qpe}}$ in the 2014 data}\label{app:T_qpe}

\begin{figure}[h!]
	\centering
    \includegraphics[width=\columnwidth]{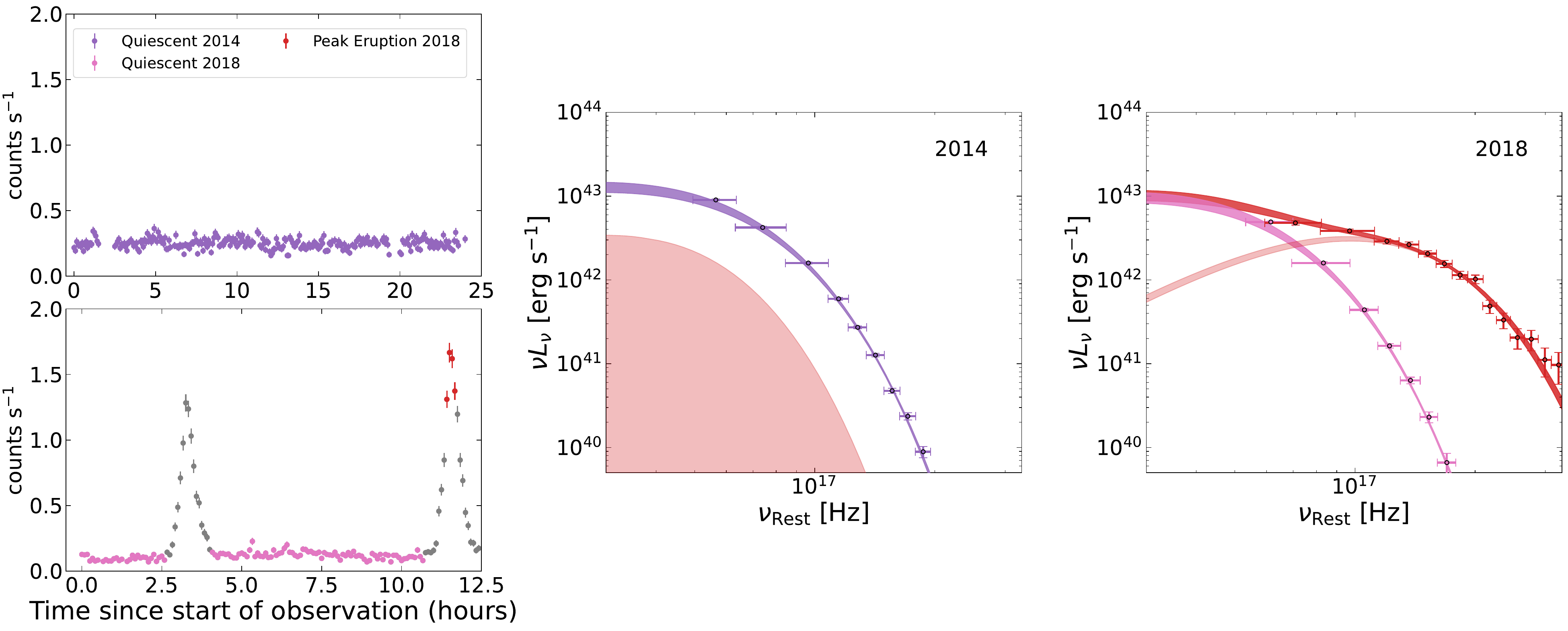}\\
    \vspace{1cm}
	\includegraphics[width=\columnwidth]{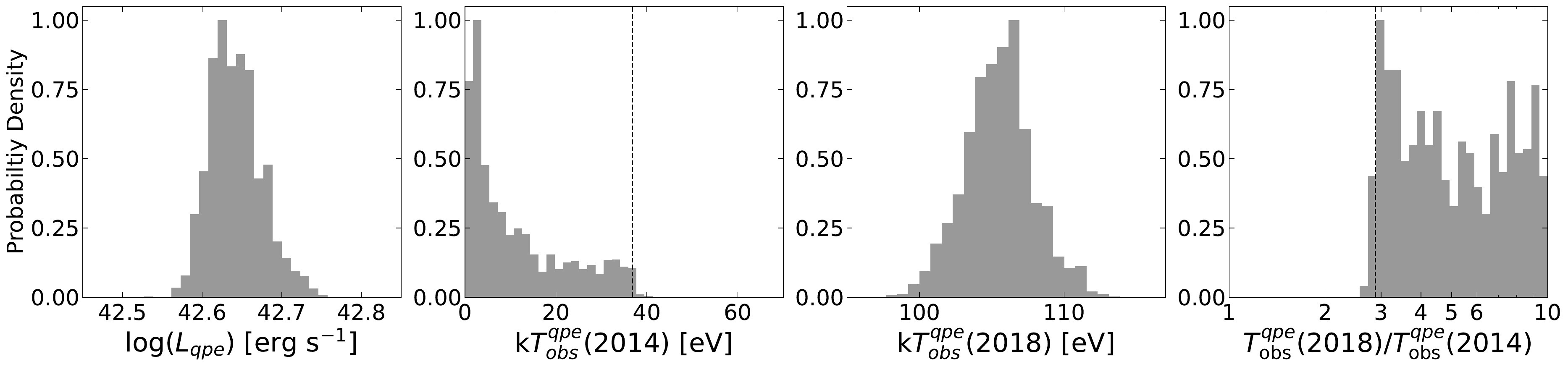}
 
	\caption{Simultaneous fitting of the disk and eruption components in 2014 and 2018. \textbf{Top:} The left panel shows the intra-observation \xmm\ light curves, with colors indicating the intervals used to extract the corresponding spectra. The middle and right panels show the best-fit intrinsic (i.e., absorption-corrected) spectra: the purple/pink curves represent the disk component, while the red curves represent the blackbody/eruption component. Contours denote 68\% credible intervals, except for the red contours in the middle panel, which show the 99\% credible interval for the (undetected) eruption component in 2014. \textbf{Bottom:} Posterior probability densities for the free parameters of the eruption component. In the panels showing the 2014 temperature, the dotted lines marks the 99\% credible limit.}

    \label{fig:Tqpe14}
\end{figure}

When testing orbiter–disk collision models for QPEs, particularly the framework proposed by \citet{Linial2023}, a crucial ingredient for assessing the model’s feasibility in explaining the lack of eruptions in 2014 is the maximum observable eruption temperature at that epoch, $T_{\rm obs}^{\rm qpe}$. In their model, this temperature is expected to scale with surface density as $T_{\rm obs}^{\rm qpe} \propto \Sigma(r_{\rm orb})^{-5/2}$. To constrain this value, we extend our disk-fitting approach by introducing an additional dataset—namely, the spectrum at the peak of the 2018 eruption (see red in Fig.~\ref{fig:Tqpe14})—as well as a new model component (\texttt{blackbody} in \texttt{XSPEC}) to represent the additional spectral component associated with the eruptions.

We also apply this component to the 2014 spectrum. According to the \citet{Linial2023} model, eruptions should still occur at this epoch, but their temperatures would be sufficiently low to evade detection in the X-rays. To fit the free parameters, $T_{\rm obs}^{\rm qpe}$ and $L_{\rm qpe}$ (the bolometric luminosity of the eruption component), we make use of a key prediction from their model: namely, that $L_{\rm qpe} \propto \Sigma(r_{\rm orb})^0$, i.e., the eruption (bolometric) luminosity should remain constant between epochs.

Therefore, our fit includes three free parameters: $L_{\rm qpe}$, $T_{\rm obs}^{\rm qpe}(2014)$, and $T_{\rm obs}^{\rm qpe}(2018)$. In practice, the 2014 eruption component must be cooler than a threshold temperature such that its emission remains below the uncertainty level of the observed 2014 disk spectrum (see middle panel of Fig.~\ref{fig:Tqpe14}), otherwise they would be detectable in the 2014 light curve (left panel Fig.~\ref{fig:Tqpe14}). By simultaneously fitting the 2014 and 2018 epochs—modeling the total spectra as the sum of the disk and eruption components—we are able to place meaningful constraints on these parameters.

The bottom panels of Fig.~\ref{fig:Tqpe14} show the posterior probability densities for the eruption parameters. Of course, since the eruptions are not detected 2014, we can only obtain a maximum temperature (upper limit) on the $\mathrm{k}T_{\rm obs}^{\rm qpe}(2014)$, while a proper measurement can be made in 2018 when these were detected. For the purposes of the discussion in \S\ref{sec:qpes}, we find (99\% credible interval) that
$\mathrm{k}T_{\rm obs}^{\rm qpe}(2014) \leq 36~\mathrm{eV}$, and
$T_{\rm obs}^{\rm qpe}(2018) / T_{\rm obs}^{\rm qpe}(2014) \geq 2.9$.

\end{document}